\documentclass[11pt]{article}

\usepackage[english]{babel}
\usepackage{amsmath}
\usepackage{amssymb}

\newtheorem{theorem}{Theorem}
\newtheorem{lemma}{Lemma}
\newtheorem{proposition}{Proposition}
\newtheorem{remark}{Remark}

\newcommand{\mG}{\mu_G}
\newcommand{\LG}{L^2(G,\mu_G)}

\newcommand{\emme}{\mathsf{m}}
\newcommand{\maxi}{\bar{K}}
\newcommand{\sigmas}{\mathsf{s}}
\newcommand{\hilb}{\mathcal{G}^\chi}
\newcommand{\hilbm}{\mathcal{G}_{\mu_{G,K}}^\chi}
\newcommand{\isom}{\mathsf{I}_{\mu_{G,K}}}
\newcommand{\repr}{R^\chi_{\mu_{G,K}}}
\newcommand{\repres}{R^\chi}
\newcommand{\unop}{\hat{F}_{\mathsf{s}}}
\newcommand{\act}{\mathfrak{a}}
\newcommand{\ppp}{\mathbf{p}}
\newcommand{\qqq}{\mathbf{q}}
\newcommand{\Pn}{\mathbf{P}}
\newcommand{\Qn}{\mathbf{Q}}
\newcommand{\iso}{\delta}
\newcommand{\para}{\boldsymbol{z}}
\newcommand{\sigmab}{\breve{\mathsf{s}}}
\newcommand{\RRR}{\mathbf{R}}
\newcommand{\rrr}{\mathbf{r}}
\newcommand{\TTT}{\mathrm{T}}
\newcommand{\SSS}{\mathrm{S}}
\newcommand{\BBB}{\mathrm{B}}
\newcommand{\AAA}{\mathrm{A}}
\newcommand{\noG}{G\!\!\!\!/}
\newcommand{\noX}{X\!\!\!\!/}

\begin{document}


\title{Square integrable projective representations \\ and \\
square integrable representations modulo a relatively central subgroup
(I): basic results}
\author{Paolo Aniello  \\ {\small{\it
Dipartimento di Scienze Fisiche,
Universit\`a di Napoli `Federico II', Napoli, Italy}}
\\{\small and} \\
{\small {\it Istituto Nazionale di Fisica Nucleare,
Sezione di Napoli, Italy}}}


\maketitle


\begin{abstract}
We introduce the notion of square integrable group representation
modulo a relatively central subgroup and, establishing a link with
square integrable projective representations,
we prove a generalization of
a classical theorem of Duflo and Moore. As an example, we apply the results
obtained to the Weyl-Heisenberg group.
\end{abstract}


\section{Introduction}

Square integrable representations of locally compact groups
have important applications in many fields
of physics (generalized coherent states, quantization,
quantum measurement theory, signal analysis etc.; see the review
paper~\cite{Ali}, the recent book~\cite{Ali2} and the rich bibliography
therein) and mathematics (the theory of Plancherel measure
for locally compact groups~\cite{Duflo},
wavelet analysis~\cite{Daubechies},
its generalization and the theory of localization operators~\cite{Wong}
etc.).\\
The fundamental properties of these representations have been studied
originally by Godement, in the case of unimodular
groups~\cite{Godement}~\cite{Godement*}, and by
Duflo and Moore~\cite{Duflo},
Phillips~\cite{Phillips}, Carey~\cite{Carey},
Grossmann et al.~\cite{Grossmann}, in the general case.
The notion of square integrable representation, modulo a central subgroup,
of a unimodular group has been studied by A.\ Borel~\cite{Borel}.

In the present paper, we introduce the notion of square-integrability,
modulo a relatively central subgroup, of a representation, which extends
the notion of square-integrability modulo a central subgroup, thus,
in particular, the simple square-integrability. Then, we show that
the square-integrability of a representation of a locally compact
group $G$, modulo a relatively central subgroup $K$ (which is a
normal subgroup of $G$), is equivalent to the square-integrability
of a projective representation of the quotient group $X=G/K$, hence,
to the square-integrability of a unitary representation of a
central extension of the circle group $\mathbb{T}$ by $X$.
This procedure allows to prove a generalization of the already cited
classical result of Duflo and Moore. In the meantime, it is {\it operative},
in the sense that it can be directly applied to concrete cases, as we
show for the representations of the Weyl-Heisenberg group.
This example is remarkable since
it is related to the classical coherent states of Schr\"odinger~\cite{Schr},
Glauber~\cite{Glauber}, Klauder~\cite{Klauder} and
Sudarshan~\cite{Sudarshan}.
More examples and applications will be given in a companion
paper~\cite{Aniello*}.

The paper is organized as follows. In section~{\ref{2}}, we review the
main properties of square integrable unitary representations, in particular
the classical theorem of Duflo and Moore. In section~{\ref{3}}, we introduce
the notion of square integrable projective representation and prove the,
so to say, `Duflo-Moore theorem for projective representations'. Next,
in section~{\ref{4}}, we define the notion of square integrable
representation modulo a relatively central subgroup and, using the
results of sections~{\ref{2}} and~{\ref{3}}, we prove a generalization
of the theorem of Duflo and Moore and other basic results.
Then, in section~{\ref{5}}, we study the intertwining properties
associated with square integrable representations modulo a relatively central
subgroup. Eventually, in section~{\ref{6}}, we discuss the main results
obtained and we apply them to the representations of the
Weyl-Heisenberg group and to another interesting example.


\section{Square integrable unitary representations} \label{2}

Let $G$ be a locally compact second countable Hausdorff topological group
(in short, l.c.s.c.\ group). We will denote by $\mG$ 
a {\it left Haar measure} (of course uniquely defined up to multiplication
by a positive constant\footnote{In order to stress the essential unicity
of the Haar measure, we will often call a particular choice of this
measure a {\it normalization} of the Haar measure.}) on $G$ and by $\Delta_G$ the modular
function on $G$.
We recall that the the {\it left regular representation} $R$
of $G$ in $\LG$ is the strongly continuous unitary representation
defined by
\begin{equation}
\left(R_g f\right)(g^\prime) = f(g^{-1}g^\prime),\ \ \
g,g^\prime\in G,
\end{equation}
for all $f\in\LG$.

Let $\mathcal{H}$ be a separable complex Hilbert space.
We will denote by $\langle\,\cdot\, ,\cdot\,\rangle$
the inner product in $\mathcal{H}$, which we will assume to be
linear in the {\it second} argument, and by $\|\cdot\|$ the
associated norm.
We will say that a linear operator $C$ from $\mathcal{H}$ into
a complex Hilbert space $\mathcal{H}^\prime$ is {\it essentially isometric}
if it is a multiple of an isometry, i.e.\ if there exists an isometry
$\mathcal{J} :\mathcal{H}\rightarrow\mathcal{H}^\prime$ such that
$C=\lambda\,\mathcal{J}$, with $\lambda >0$.

Let $U$ be a strongly continuous irreducible unitary representation of $G$
in $\mathcal{H}$. Given a couple of vectors
$\phi,\psi\in\mathcal{H}$, we can define the `coefficient' 
\begin{equation}
c_{\psi ,\phi}^U :\ G\ni g\mapsto
\langle U(g) \psi , \phi \rangle\in\mathbb{C},
\end{equation}
which is a bounded continuous function,
and the set (of `admissible vectors for $U$')
\begin{equation}
\mathcal{A}(U):=\left\{\psi\in\mathcal{H}\,|\ \exists\phi\in\mathcal{H}:\,
\phi\neq 0,\, c_{\psi,\phi}^U \in L^2(G,\mu_G)\right\}.
\end{equation}
Then, the representation $U$ is said to be {\it square integrable} if
\[
\mathcal{A}(U)\neq\{0\}.
\]
Since $U$ is irreducible, this condition is equivalent to the
existence of a square integrable nonzero coefficient $c_{\psi,\phi}^U$.\\
Square integrable representations are described by 
the following classical result due to Duflo and Moore
(see \cite{Duflo}).

\begin{theorem} \label{Duflo}
Let the strongly continuous irreducible unitary representation
$U$ of the l.c.s.c.\ group $G$ in the Hilbert space $\mathcal{H}$
be square integrable. Then, the set $\mathcal{A}(U)$
is a dense linear manifold in $\mathcal{H}$ and,
for any couple of vectors
$\phi\in\mathcal{H}$
and $\psi\in\mathcal{A}(U)$, the
cofficient $c_{\psi,\phi}^U$ is square integrable with respect to
the left Haar measure $\mu_G$ on $G$. Moreover, for any
nonzero $\psi\in\mathcal{A}(U)$, the map
\begin{equation}
C_{\psi}^U :\ \mathcal{H}\ni\phi\mapsto c_{\psi,\phi}^U\in\LG
\end{equation}
defines a linear operator which is essentially isometric
and intertwines $U$ with the left regular representation
of $G$ in $\LG$, namely
\begin{equation}
C_{\psi}^U \, U(g) = R_g \, C_{\psi}^U,\ \ \
\forall g\in G.
\end{equation}
Finally, there exists a unique positive selfadjoint injective linear
operator $D_U$ in $\mathcal{H}$, such that
\[
\mathcal{A}(U)=\mathrm{Dom}\left(D_U\right) 
\]
and
\begin{eqnarray} \label{ortho}
\int_G c_{\psi_1,\phi_1}^U(g)^{*}\,c_{\psi_2,\phi_2}^U(g)\ d\mu_G(g)
& \!\! = \!\! & 
\int_G \langle\phi_1,U(g)\,\psi_1\rangle\, \langle U(g)\,\psi_2,\phi_2
\rangle\ d\mu_G (g)\nonumber\\
& \!\! = \!\! &
\langle\phi_1 ,\phi_2\rangle\,
\langle D_U\,\psi_2, D_U\,\psi_1\rangle ,
\end{eqnarray}
for all $\phi_1,\phi_2\in\mathcal{H}$, for all $\psi_1,\psi_2\in
\mathcal{A}(U)$. The operator $D_U$ is bounded if and only if
$G$ is unimodular and, in such case, it is a multiple of the identity.
\end{theorem}

If $U$ is square integrable,
the operator $D_U$ of Theorem~{\ref{Duflo}} --- which we will call the
{\it Duflo-Moore operator} --- being injective and selfadjoint, has
a densely defined selfadjoint inverse $D_U^{-1}$ (see, for
instance,~\cite{Rudin}, Theorem~{13.11}).
Duflo and Moore call
the square of $D_U^{-1}$ the formal degree of the representation $U$.
Notice that the operator $D_U$ depends on the normalization of the
Haar measure $\mu_G$. Indeed, if $\mu_G$ is rescaled by a positive
constant, then $D_U$ is rescaled by the square root of the same constant.
Thus, we will say that $D_U$ is {\it normalized according to} $\mu_G$.\\
The theorem of Duflo and Moore has some important implications.
Let us list the main ones.
\begin{enumerate}

\item The square-integrability of a unitary representation
depends only on its
unitary equivalence class.
According to Theorem~{\ref{Duflo}}, if
$U$ is square integrable, then it is unitarily equivalent to a
subrepresentation $U_R$ of the left regular representation $R$.

\item Let $G$ be a compact group. Then, any strongly continuous irreducible
unitary representation of $G$
is square integrable. This follows from the fact that, in this case,
the Haar measure on $G$  is finite. Moreover, in this case $G$ is unimodular
so that the Duflo-Moore operators associated with its irreducible
unitary representations are simply multiples of the identity.
In fact, for a compact group, Theorem~{\ref{Duflo}} reduces to
a well known classical result (see, for instance, \cite{Gaal}).

\item If the representation $U$ of $G$ is square integrable, then,
according to Theorem~{\ref{Duflo}}, for any nonzero
admissible vector $\psi\in\mathcal{A}(U)$, one can define the
linear operator
\begin{equation}
W_{\psi}^U:\ \mathcal{H}\ni \phi\mapsto
\| D_U\,\psi\|^{-1}\,
c_{\psi ,\phi}^U \in L^2(G,\mu_G)
\end{equation}
--- sometimes called {\it generalized wavelet transform}
generated by $U$, with {\it analyzing vector} $\psi$ ---
which is an isometry. The ordinary wavelet transform arises as
a special case when $G$ is the (1+1)-dimensional affine group,
i.e.\ the semidirect product $\mathbb{R}\times^\prime\mathbb{R}_*^+$
(see \cite{Grossmann} and \cite{Grossmann*}).

\item The range $\mathcal{R}_\psi^U$ of $W_{\psi}^U$
(or $C_\psi^U$), which consists of
bounded continuous functions, is
a reproducing kernel Hilbert space
(a classical reference on r.k.H.s.\ is~{\cite{Arons}})
and the reproducing kernel is given explicitly by:
\begin{equation}
\varkappa_\psi^U(g,g^\prime)= \| D_U\,\psi\|^{-2}\,
\langle U(g)\,\psi ,U(g^\prime)\, \psi\rangle,
\ \ g,g^\prime\in G.
\end{equation}
Namely, for any function $f$ in $\mathcal{R}_\psi^U$, we have:
\begin{equation}
f(g)=\int_G \varkappa_\psi^U(g,g^\prime)\, f(g^\prime )\ d\mu_G(g^\prime),\ \
\forall g\in G.
\end{equation}
This property follows from the `orthogonality relation'~{(\ref{ortho})}.

\end{enumerate}

Let us prove an interesting invariance property of the Duflo-Moore
operator with respect to the representation $U$ (Duflo and Moore used
a similar property of the formal degree operator for proving their
classical result).

\begin{proposition} \label{semi}
Let $U$ be square integrable.
Then, the dense linear manifold
$\mathrm{Dom}(D_U)=\mathcal{A}(U)$ is invariant with respect to
$U$ and the positive selfadjoint operator $D_U$ is
semi-invariant with weight $\Delta_G^{1/2}$, i.e.
\begin{equation}
U(g)\, D_U\, U(g)^{-1}=\Delta_G(g)^{1/2}\, D_U,\ \ \ \
\forall g\in G.
\end{equation}
\end{proposition}

{\bf Proof\, :} The linear manifold $\mathcal{A}(U)$ is invariant with respect
to $U$; indeed:
\[
\int_G |\langle U(g^\prime)\,U(g)\,\psi,\phi\rangle|^2\ d\mu_G(g^\prime)=
\Delta_G(g)^{-1} \int_G |\langle U(g^\prime)\,\psi,
\phi\rangle |^2 d\mu_G(g^\prime).
\]
Now, let $U$ be square integrable. Then, given $\phi\in\mathcal{H}$,
$\|\phi\|=1$, for any $\psi_1,\psi_2\in\mathrm{Dom}(D_U)$, we have:
\begin{eqnarray*}
\langle D_U\,U(g^{-1}) \psi_2,D_U\,\psi_1  \rangle
\!\! & = & \!\!
\int_G \langle \phi,U(g^\prime)\,\psi_1\rangle
\langle U(g^\prime g^{-1})\,\psi_2,\phi\rangle\ d\mu_G(g^\prime)
\\
\!\! & = &  \!\!
\Delta_G(g)\,
\int_G \langle\phi,U(g^\prime g)\,\psi_1\rangle \langle U(g^\prime)\,\psi_2,
\phi\rangle\ d\mu_G(g^\prime)
\\
\!\! & = & \!\!
\Delta_G(g)\,\langle D_U\,\psi_2, D_U\,U(g)\,\psi_1  \rangle.
\end{eqnarray*}
Since $D_U$ is a densely defined selfadjoint operator,
$D_U^2$ is a densely defined positive selfadjoint operator (whose domain
is a core for $D_U$).
If $\psi_1$ belongs to $\mathrm{Dom}(D_U^2)$, we obtain:
\[
\langle \psi_2,U(g)\,D_U^2\,\psi_1\rangle=
\Delta_G(g)\,\langle D_U\,\psi_2, D_U\,U(g)\,\psi_1\rangle,\ \ \
\forall\psi_2\in\mathrm{Dom}(D_U).
\]
From this relation, since $D_U$ is selfadjoint, it follows that
\[
D_U U(g)\,\psi_1\in\mathrm{Dom}(D_U)\ \ 
\mbox{and}\ \ 
\langle\psi_2,U(g)\,D_U^2\,\psi_1\rangle=
\Delta_G(g)\,\langle\psi_2,D_U^2\, U(g)\,\psi_1\rangle.
\]
Thus, the domain of $D_U^2$ is invariant with respect to $U$ and,
by the arbitrariness of $\psi_2$ in the dense domain of
$D_U$, we deduce that
\[
\left(U(g)\,D_U U(g)^{-1}\right)\left(U(g)\,D_U U(g)^{-1}\right)=
U(g)\,D_U^2\,U(g)^{-1}=\Delta_G(g)\, D_U^2,
\]
for all $g\in G$. Eventually,
since the square root of a positive selfadjoint operator
is unique, $D_U$ is semi-invariant with weight $\Delta_G^{1/2}$ and
the proof is complete.~$\blacksquare$

In many physical applications, one has to deal with
representations that are more general than unitary representations,
namely with {\it projective representations}.
Thus, in the next section, we will extend the notion of
square-integrability to projective representations.
This will also allow us to prove, in section~{\ref{4}}, the main
results of this paper.


\section{Square integrable projective representations}
\label{3}

Let $P$ be a {\it projective representation} of a l.c.s.c.\ group $G$ in
a separable complex Hilbert space $\mathcal{H}$
(see, for intance, \cite{Raja}, chapter~{VII}),
namely a map of $G$ into $\mathcal{U}(\mathcal{H})$, the unitary group of
$\mathcal{H}$, such that
\begin{description}
\item[\em{\small 1)}]\ \ \ 
$P$ is a weakly Borel map, i.e.\
$G\ni g\mapsto \langle\phi,P(g)\,\psi\rangle\in\mathbb{C}$ is a Borel
function\footnote{The terms {\it Borel function} (or map)
and {\it Borel measure} will be always used with reference to
the natural Borel strucures on the topological spaces involved,
namely to the smallest $\sigma$-algebras containing all open subsets.},
for any $\phi,\psi\in\mathcal{H}$;
\item[\em{\small 2)}]\ \ \ 
$P(e)=I$, where $e$ is the identity in $G$ and
$I$ the identity operator;
\item[\em{\small 3)}]\ \ \ 
denoted by $\mathbb{T}$ the circle group, namely
the group of complex numbers of modulus one,
there exists a Borel function
$\emme : G\times G\rightarrow\mathbb{T}$ such that
\[
P(gh)=\emme (g,h)\,P(g)\,P(h),\ \ \ \forall\, g,h\in G.
\]
\end{description}
The function $\emme$, which
is called the {\it multiplier associated with} $P$,
satisfies the following conditions:
\begin{equation}
\emme(g,e)=\emme(e,g)=1,\ \ \ \ \forall g\in G,
\end{equation}
and
\begin{equation}
\emme(g_1,g_2g_3)\, \emme(g_2,g_3)=
\emme(g_1 g_2,g_3)\, \emme(g_1,g_2),\ \ \ \ \forall\, g_1,g_2,g_3\in G.
\end{equation}
In general, a Borel map $\emme: G\times G\rightarrow\mathbb{T}$
satisfying the previous conditions is said to be a multiplier for $G$
(if $\mathbb{T}$ is replaced by another abelian group $\mathbb{A}$,
$\emme$ is said to be a $\mathbb{A}$-multiplier).
Two multipliers $\emme,\emme^\prime$ for $G$ are said to be similar
if there exists a Borel function $\beta : G\rightarrow\mathbb{T}$
such that
\begin{equation}
\emme(g_1,g_2)= \beta(g_1 g_2)\,\beta(g_1)^{-1}\beta(g_2)^{-1}
\,\emme^\prime(g_1,g_2),\ \ \ \ \forall\, g_1,g_2\in G.
\end{equation}

Irreducibility for projective representations is defined as for
standard representations. Equivalence of projective representations is
defined as follows. Let us identify the circle group $\mathbb{T}$ with
the set $\{zI\,|\ z\in\mathbb{T}\}\subset\mathcal{U}(\mathcal{H})$ which is
the centre of $\mathcal{U}(\mathcal{H})$. Let us denote by $\varpi$
the canonical projection homomorphism of $\mathcal{U}(\mathcal{H})$ onto
$\mathcal{P}(\mathcal{H}):=\mathcal{U}(\mathcal{H})/\mathbb{T}$,
the projective group of $\mathcal{H}$. Then two projective representations
$P, Q$ of $G$ are said to be equivalent if there is a projective
representation $P^\prime$ of $G$, unitarily or antiunitarily
equivalent to $Q$, such that
\begin{equation} \label{ray}
\varpi(P(g))=\varpi(P^\prime(g)),\ \ \ \  \forall g\in G.
\end{equation}
This definition of (physical) equivalence is consistent with Wigner's theorem
on simmetry transformations~\cite{Wigner}.
Two projective representations
$P,P^\prime$ verifying relation~{(\ref{ray})} are said to be ray
equivalent. Two ray equivalent representations have similar multipliers;
conversely, if $P$ is a projective representation of $G$
with multiplier $\emme$
and $\emme^\prime$ is a multiplier similar to $\emme$, then there
exists a projective representation $P^\prime$ of $G$, ray equivalent to
$P$, with multiplier $\emme^\prime$. 

Now, given the cartesian product $\mathbb{T}\times G$, the composition law
\begin{equation}
(\tau,g)(\tau^\prime,h)=(\emme(g,h)\tau\tau^\prime, gh)
\end{equation}
defines a group $G_\emme$.
It is well known that
there exists a unique topology on $\mathbb{T}\times G$
that makes $G_\emme$ a l.c.s.c.\ topological group and generates a
Borel structure on $G_\emme$ which coincides with the product Borel structure
on $\mathbb{T}\times G$. The group $G_\emme$ is a central extension of
$\mathbb{T}$ by $G$. One can check easily that
a left Haar measure on  $G_\emme$ is given
by the product measure $\mu_\mathbb{T}\otimes\mG$, where $\mu_\mathbb{T}$
is the Haar measure on $\mathbb{T}$ (as usual for compact groups,
we will assume that $\mu_\mathbb{T}(\mathbb{T})=1)$, and the modular
function on $G_\emme$ is given by
\begin{equation} \label{modular}
\Delta_{G_\emme}(\tau, g)=\Delta_G(g),\ \ \ \ \forall\tau\in\mathbb{T},\
\forall g\in G;
\end{equation}
hence, $G_\emme$ is unimodular if and
only if $G$ is. If $\emme^\prime$ is a multiplier for $G$ similar to
$\emme$, then $G_{\emme^\prime}$ is isomorphic, as a topological
group, to $G_\emme$. \\
The map
\begin{equation}
U_P:\
G_\emme\ni (\tau,g) \mapsto \tau^{-1} P(g)\in\mathcal{U}(\mathcal{H})
\end{equation}
is a unitary representation of $G_\emme$ in $\mathcal{H}$ which is weakly
Borel, hence, according to a classical result (see, for instance,
\cite{Raja}),
strongly continuous. It is trivial to show that $U_P$ is
irreducible if and only if $P$ is. One can check that the multiplier $\emme$
defines a projective representation of $G$ in $L^2(G,\mG)$,
with multiplier $\emme$, by
\begin{equation}
\left(R^{\emme}_g\, f\right)(g^\prime)=
\emme(g,g^{-1} g^\prime)^{-1}\,
f(g^{-1} g^\prime),\ \ \ f\in L^2(G,\mG).
\end{equation}
We will call $R^\emme$ the {\it left regular $\emme$-representation} of
$G$.

Obviously, given a couple of vectors in $\mathcal{H}$,
one can define a coefficient
function associated with $P$ precisely in the same way as it has been
done for a unitary representation. Then, one can
define the set of admissible vectors for $P$, i.e.
\begin{equation}
\mathcal{A}(P):=\left\{\psi\in\mathcal{H}\,|\ \exists\phi\in\mathcal{H}:\,
\phi\neq 0,\, c_{\psi,\phi}^P \in L^2(G,\mu_G)\right\}.
\end{equation}
At this point, if $P$ is irreducible, one
says that $P$ is {\it square integrable} if
$\mathcal{A}(U)\neq\{0\}$. Thus, for unitary representations this
definition coincides with the one given in section~{\ref{2}}.
Let us show that square integrable projective representations enjoy
properties analogous to that of square integrable unitary representations.

\begin{theorem} \label{Duflo2}

Let $P$ be an irreducible projective representation
of the l.c.s.c.\ group $G$ in the Hilbert space $\mathcal{H}$ and
$\emme$ the associated multiplier. Then, $P$ is square integrable if and
only if $U_P$ is a square integrable unitary representation of $G_\emme$ in
$\mathcal{H}$. Moreover, any projective representation of
$G$ equivalent to $P$ is square integrable if and only $P$ is.
\\
Assume that $P$ is square integrable.
Then, $\mathcal{A}(P)$ is a dense linear manifold in $\mathcal{H}$ and
$\mathcal{A}(P)=\mathcal{A}(U_P)$.
For any $\phi\in\mathcal{H}$
and any $\psi\in\mathcal{A}(P)$, the function
\begin{equation}
c_{\psi,\phi}^P:\
G\ni g\mapsto\langle P_g\,\psi ,\phi\rangle\in\mathbb{C}
\end{equation}
is square integrable with respect to the left Haar measure
$\mu_G$ on $G$ and, if $\psi\neq 0$, the map
\begin{equation}
C_\psi^P :\ \mathcal{H}\ni\phi\mapsto c_{\psi,\phi}^P\in
L^2(G,\mG)
\end{equation}
defines a linear operator which is essentially isometric and intertwines
$P$ with the left regular $\emme$-representation of $G$; namely:
\begin{equation} \label{inter}
C_\psi^P \, P(g) = R_g^\emme \, C_\psi^P,\ \ \ \forall g\in G.
\end{equation}
There exists a unique positive selfadjoint injective
linear operator $D_P$ in $\mathcal{H}$ such that
\[
\mathcal{A}(P)=\mathrm{Dom}\left(D_P\right) 
\]
and
\begin{equation} 
\int_G c_{\psi_1,\phi_1}^P (g)^{*}\,c_{\psi_2,\phi_2}^P (g)\ d\mu_G(g) =
\langle\phi_1 ,\phi_2\rangle\,
\langle D_P\,\psi_2, D_P\,\psi_1\rangle ,
\end{equation}
for all $\phi_1,\phi_2\in\mathcal{H}$, for all $\psi_1,\psi_2\in
\mathcal{A}(P)$. Moreover, $D_P$ is equal to the
Duflo-Moore operator $D_{U_P}$
associated with the square integrable representation $U_P$, provided
that $D_{U_P}$ is normalized according to $\mu_G\!\otimes\!\mu_\mathbb{T}$,
with $\mu_\mathbb{T}(\mathbb{T})=1$.
Finally, $D_P$ is bounded if and only if
$G$ is unimodular and, in such case, it is a multiple of the identity.

\end{theorem}

{\bf Proof\,:} The map
\[
\mathbb{T}\times G\ni(\tau,g)\mapsto |\langle P(g)\,\psi,
\phi\rangle|^2
\]
is a non-negative Borel function; hence:
\begin{eqnarray*}
\int_{G_\emme}\! |\langle U_P(\tau, g)\,\psi,\phi\rangle|^2\
d \mu_\mathbb{T}\!\otimes\!\mu_G(\tau,g)
& \!\! = \!\! &
\int_{\mathbb{T}\times G}\! |\langle P(g)\,\psi,\phi\rangle|^2\
d \mu_\mathbb{T}\!\otimes\!\mu_G(\tau,g)
\\
\mbox{(Tonelli's theorem)}
\!\! & = & \!\!
\int_G |\langle P(g)\psi,\phi\rangle|^2\
d\mu_G(g) \int_\mathbb{T}\! d\mu_\mathbb{T}(\tau)
\\
(\,\mu_\mathbb{T}=1\,) 
\!\! & = & \!\!
\int_G |\langle P(g)\psi,\phi\rangle|^2\ d\mu_G(g).
\end{eqnarray*}
Thus, we have that $\mathcal{A}(P)=\mathcal{A}(U_P)$ and
$P$ is square integrable if and only if $U_P$ is.
Moreover,
if $P^\prime$ is a projective representation of $G$ equivalent to $P$,
there exist a unitary or antiunitary operator $V$ and a Borel function
$\epsilon : G\rightarrow\mathbb{T}$ such that
\[
P^\prime(g)=\epsilon(g)\, V^* P(g)\, V,\ \ \ \ \forall g\in G,
\]
and hence:
\[
|c_{\psi,\phi}^P(g)|=|c_{V\psi,V\phi}^{P^\prime}(g)|,\ \ \ \ \forall g\in G.
\]
It follows that $P^\prime$ is square integrable if and only if $P$ is.\\
Let $P$ be square integrable. Then, $U_P$ is square integrable and
we have already shown that $\mathcal{A}(P)=\mathcal{A}(U_P)$.
Now, notice that
\[
c_{\psi_1,\phi_1}^{U_P}(\tau, g)^*\, c_{\psi_2,\phi_2}^{U_P}(\tau, g)=
c_{\psi_1,\phi_1}^{P}(g)^*\, c_{\psi_2,\phi_2}^{P}(g),\ \ \ \
\forall\tau\in\mathbb{T},\ \forall g\in G.
\]
Thus, according to the theorem of Duflo and Moore, there is a unique
positive selfadjoint injective operator $D_{U_P}$
such that $\mathcal{A}(U_P)=\mathrm{Dom}(D_{U_P})$ and,
for any $\phi_1,\phi_2
\in\mathcal{H}$, $\psi_1,\psi_2\in\mathcal{A}(U_P)=\mathcal{A}(P)$,
\begin{eqnarray*}
\langle \phi_1,\phi_2\rangle \langle D_{U_P}\psi_2, D_{U_P}\psi_1\rangle
\!\! & = & \!\!
\int_{G_\emme}\! c_{\psi_1\phi_1}^{U_P}(\tau, g)^*\,
c_{\psi_2,\phi_2}^{U_P}(\tau, g)\
d \mu_\mathbb{T}\!\otimes\mu_G(\tau, g)
\\
\!\! & = & \!\!
\int_G c_{\psi_1,\phi_1}^{P}(g)^*\, c_{\psi_2,\phi_2}^{P}(g)\
d\mu_G(g).
\end{eqnarray*}
Hence, the linear operator $C_\psi^P$ is essentially isometric.
Moreover, since
$G$ is unimodular if and only if $G_\emme$ is, $D_P\equiv D_{U_P}$ is
bounded if and only if $G$ is unimodular and, in such case, it is
a multiple of the identity. Finally, let us prove the
intertwining property~{(\ref{inter})}. In fact, we have:
\begin{eqnarray*}
\left(C_\psi^{P} (P(g)\,\phi)\right)(g^\prime)
\!\! & = & \!\!
\langle P(g^\prime)\,\psi,P(g)\,\phi\rangle
\\
\!\! & = & \!\!
\langle P(g)^{-1} P(g^\prime)\,\psi,\phi\rangle
\\
\!\! & = & \!\!
\emme(g,g^{-1})^{-1}\,\langle P(g^{-1})\,P(g^\prime)\,\psi,\phi\rangle
\\
\!\! & = & \!\!
\emme(g,g^{-1})^{-1}\,\emme(g^{-1},g^\prime)\,
\langle P(g^{-1}g^\prime)\,\psi,\phi\rangle
\\
\!\! & = & \!\!
\emme(g,g^{-1}g^\prime)^{-1}\,\langle P(g^{-1}g^\prime)\,\psi,\phi\rangle
\\
\!\! & = & \!\!
\left(R_g^\emme (C_\psi^P\,\phi)\right)(g^\prime) ,\ \ \ \
\psi\in\mathcal{A}(P),\ \psi\neq 0.
\end{eqnarray*}
The proof is complete.~$\blacksquare$

\begin{remark} \label{alternative}
If $\emme$ is a multiplier for $G$, then also 
\begin{equation}
\emme^*:G\ni (g,h)\mapsto \emme(g,h)^*=\emme(g,h)^{-1}\in\mathbb{T}
\end{equation}
is a multiplier and one can define the l.c.s.c.\ group $G_{\emme^*}$.
Furthermore, if $P$ is an irreducible projective representation of $G$
with multiplier $\emme$, the map
\begin{equation} \label{star}
U_{*P}: G_{\emme^*}\ni(\tau,g)\mapsto \tau\, P(g)\in\mathcal{U}(\mathcal{H})
\end{equation}
is a strongly continuous irreducible unitary representation of
$G_{\emme^*}$. Then, arguing as above,
one can substitute in Theorem~\ref{Duflo2}
the representation $U_P$ with $U_{*P}$.
\end{remark}

\begin{remark}
One can show easily that the linear manifold $\mathcal{A}(P)$ is invariant
with respect to $P$. If $P$ is square integrable, then, according to
Proposition~{\ref{semi}}, the Duflo-Moore operator $D_{U_P}$ associated with
the square integrable unitary representation $U_P$ is
semi-invariant with weight $\Delta_{G_\emme}$; hence, recalling
equation~{(\ref{modular})}, since $P(g)=U_P(1,g)$ and $D_P=D_{U_P}$,
we have:
\[
P(g)\, D_P P(g)^{-1} = \Delta_G(g)^{1/2}\, D_P,\ \ \ \ \forall g\in G.
\]
\end{remark}

Notice that, as it happens for square integrable unitary representations,
if one rescales the Haar measure
$\mu_G$ by a positive constant, the operator $D_P$ is rescaled by
the square root of the same constant.
We will say, then, that $D_P$ is {\it normalized according to} $\mu_G$.


\section{Square integrable representations modulo a relatively central
subgroup}
\label{4}

Let $G$ be a l.c.s.c.\ group, $U$ a strongly continuous irreducible unitary
representation of $G$ in a separable complex Hilbert space $\mathcal{H}$
and $K$ a closed normal subgroup of $G$
such that the restriction of $U$ to K is a scalar
representation; namely:
\begin{equation}
U(kg)=\chi(k)\, U(g)=U(gk),\ \ \ \ \forall k\in K,\ \forall g\in G,
\end{equation}
where $\chi:K\rightarrow\mathbb{T}$ is a continuous
group homomorphism.
We will say that the subgroup $K$, with the specified properties, is
$U$-{\it central} or that $K$ is {\it relatively central}
with respect to $U$. This terminology refers to the fact that $U(K)$
is a subgroup of the centre of $\mathcal{U}(\mathcal{H})$ which,
as we have seen, can be identified with $\mathbb{T}$.
For instance, any closed central subgroup $K$ of $G$ is $U$-central;
indeed, in such case we have that
\[
U(k)\, U(g)=U(kg)=U(gk)=U(g)\, U(k),\ \ \ \ \forall k\in K,\
\forall g\in G,
\]
hence, by Schur's lemma, for any $k\in K$,
$U(k)=\chi(k)\, I$, where $\chi$ is a unitary character
of the abelian group $K$. In particular,
we will denote by $K_0$ the centre of $G$. There exists a unique
maximal $U$-central subgroup of $G$ which will be denoted by
$\maxi$. It coincides with the kernel of the continuous group homomorphism
$G\ni g\mapsto\varpi\circ U(g)\in\mathcal{P}(\mathcal{H})$.


Given a generic relatively central subgroup $K$ of $G$,
we will denote by $X$ the left coset space $G/K$, which, endowed with
the quotient group structure and the quotient topology, is a l.c.s.c.\
group since $K$ is a closed normal subgroup, and by
$\mathsf{p}:G\rightarrow X$ the canonical projection homomorphism,
\begin{equation}
\mathsf{p}(g)=g\, K,\ \ \ \ g\in G,
\end{equation}
which is an open continuous map.
There is a natural continuous action of $G$ on $X$,
$(\cdot)[\,\cdot\,] : G\times X\rightarrow X$,
defined by:
\begin{equation}
g[x]=\mathsf{p}(g)\,x,\ \ \ \ g\in G,\ x\in X.
\end{equation}
Notice that a left Haar measure $\mu_X$ on $X$ is invariant with repect
to this action. Hence, it is a standard result that, denoted by
$\Delta_K$ the modular function of $K$, the following relation holds:
\[
\Delta_K(k)=\Delta_G(k),\ \ \ \ \forall k\in K.
\]
This implies that, if $G$ is unimodular, any closed normal subgroup of $G$
--- in particular, any $U$-central subgroup --- will be unimodular.
We recall also that there exists a (in general not unique) Borel map
$\sigmas : X\rightarrow G$, such that
\[
\mathsf{p}(\sigmas(x))=x, \ \  \forall x\in X,\ \ \ \mbox{and}\ \ \
\sigmas(e\,K)=e.
\]
Such a map is said to be a {\it Borel section}. Then its range
intersects each left $K$-coset in exactly one point.
Now, if $\sigmas$ is a Borel section, since $X$ is a quotient group,
we have:
\begin{equation}
\sigmas(x_1 x_2)=\sigmas(x_1)\,\sigmas(x_2)\, \kappa_\sigmas(x_1,x_2),
\end{equation}
where $\kappa_\sigmas\, : X\times X\rightarrow K$ is Borel map (if, in
particular, $K$ is a central subgroup, one can check that
$\kappa_\sigmas$ is a $K$-multiplier).

As a first step,
we want to show that if $U$ is a square integrable representation
$\maxi$ must be compact. To this aim, we need to prove a technical result
which will be extremely useful in the following (see
formula~{(\ref{decomposition})} below).

\begin{lemma} \label{lemma}
For any Borel section $\sigmas$, the map
\begin{equation}
\gamma_\sigmas : X\times K\ni (x,k)\mapsto \sigmas(x)\, k\in G
\end{equation}
is a Borel isomorphism and the image, through $\gamma_\sigmas^{-1}$,
of the product in $G$ is given by
\begin{equation} \label{product}
(x,k)\,(x^\prime,k^\prime)=
(x x^\prime, \kappa_\sigmas(x,x^\prime)^{-1}
k_{\sigmas(x^\prime)} k^\prime),
\ \ \  x,x^\prime\in X,\ k,k^\prime\in K,
\end{equation}
where we have set
$k_{\sigmas(x^\prime)}\equiv\sigmas(x^\prime)^{-1} k\,\sigmas(x^\prime)$.
\end{lemma}

{\bf Proof\, :} Since $\sigmas$ is a Borel section, $\gamma_\sigmas$ is a
bijective Borel map, hence, as $X\times K$ and $G$ equipped with
their natural Borel structures are standard Borel spaces, a Borel
isomorphism. Besides, for any $g,g^\prime\in G$, setting
\[
(x,k)=\gamma_\sigmas^{-1}(g)=(\mathsf{p}(g),\sigmas
(\mathsf{p}(g))^{-1}g),\ \ \
(x^\prime,k^\prime)=\gamma_\sigmas^{-1}(g^\prime),
\]
we have:
\begin{eqnarray*}
g\, g^\prime
\!\! & = & \!\!
\sigmas(x)\, k\,\sigmas(x^\prime)\, k^\prime
\\
\!\! & = & \!\!
\sigmas(x)\,\sigmas(x^\prime)\,\sigmas(x^\prime)^{-1} k\,\sigmas(x^\prime)\,
k^\prime
\\
\!\! & = & \!\!
\sigmas(xx^\prime) \left(\kappa_\sigmas(x,x^\prime)^{-1} \sigmas(x^\prime)^{-1}
k\,\sigmas(x^\prime)\, k^\prime\right).
\end{eqnarray*}
Then, since $K$ is a normal subgroup, the point
$\sigmas(x^\prime)^{-1}k\,\sigmas(x^\prime)$ belongs to $K$ and
\[
\gamma_\sigmas^{-1}(gg^\prime)=(xx^\prime,\kappa_\sigmas(x,x^\prime)\,
\sigmas(x^\prime)^{-1} k\, \sigmas(x^\prime)\, k^\prime),
\]
which is precisely what we had to prove.~$\blacksquare$

At this point, it is natural to ask what is the image measure,
through the map $\gamma_\sigmas^{-1}$, of the Haar measure $\mu_G$ on $G$.

\begin{lemma} \label{main}
Denoted by $\mu_K$ a left Haar measure on $K$
and by $\mu_X$ a left Haar measure
on $X$, the image measure on $X\times K$,
through the Borel isomorphism $\gamma_\sigmas^{-1}$, of $\mu_G$
does not depend on the choice of the Borel section $\sigmas$ and is 
proportional to the product measure $\mu_X\!\otimes\!\mu_K$. Hence,
for any Borel function $f:\, G\rightarrow\mathbb{C}$
and any Borel section
$\sigmas:\, X\rightarrow G$, the following formula holds:
\begin{equation} \label{decomposition}
\int_G f(g)\ d\mu_G(g)=\int_{X\times K} f(\sigmas(x)k)\
d\mu_X\!\otimes\!\mu_K (x,k),
\end{equation} 
for a suitable normalization of the Haar measures $\mu_X$ and
$\mu_K$ which does not depend on the choice of the section $\sigmas$.
\end{lemma}

{\bf Proof\, :} Recall that
a left Haar measure on a l.c.s.c.\ group $G$ is defined
uniquely, up to multiplication by a positive constant,
by the property of being a
$\sigma$-finite left-invariant measure on the Borel $\sigma$-algebra
of $G$ (see, for instance, \cite{Raja}, chapter~V, sect.~2).
Thus, all we have to show is that
the image $\tilde{\mu}_G$, through the Borel isomorphism
$\gamma_\sigmas$, of the product measure $\mu_X\!\otimes\!\mu_K$
is a $\sigma$-finite left-invariant measure on $G$.
Indeed, $\tilde{\mu}_G$ is $\sigma$-finite since $\mu_X\!\otimes\!\mu_K$
is. Moreover, for any non-negative Borel function $f:G\rightarrow
\mathbb{R}$, we have:
\begin{eqnarray*}
\int_G f(g g^\prime)\ d\tilde{\mu}_G(g^\prime)
\!\! & = & \!\!
\int_{X\times K}
f(\sigmas(x x^\prime)\,\kappa_\sigmas(x,x^\prime)^{-1}
k_{\sigmas(x^\prime)} k^\prime)\ d\mu_X\!\otimes\!\mu_K (x^\prime,k^\prime)
\\
(\mbox{Tonelli's theorem})
\!\! & = & \!\!
\int_X\!\left(\int_K f(\sigmas(xx^\prime)\,\kappa_\sigmas(x,x^\prime)^{-1}
k_{\sigmas(x^\prime)} k^\prime)\ d\mu_K(k^\prime)\right)\! d\mu_X(x^\prime)
\\
(\mbox{invariance of $\mu_K$})
\!\! & = & \!\!
\int_K \!\left(\int_X f(\sigmas(xx^\prime)\, k^\prime)\ d\mu_X(x^\prime)
\right)\! d\mu_K(k^\prime)
\\
(\mbox{invariance of $\mu_X$})
\!\! & = & \!\!
\int_{X\times K} f(\sigmas(x^\prime)\,k^\prime)\
d\mu_X\!\otimes\!\mu_K(x^\prime,k^\prime).
\end{eqnarray*}
Hence, for any non-negative Borel function $f$
on $G$, we have
\[
\int_G f(gg^\prime)\ d\tilde{\mu}_G(g^\prime)=
\int_G f(g^\prime)\ d\tilde{\mu}_G(g^\prime),
\]
so that $\tilde{\mu}_G=\alpha\,\mu_G$, for some $\alpha>0$.
If $\sigmas^\prime$ is another Borel section, we have that
$\sigmas^\prime(x) = \sigmas(x)\,\upsilon(x)$, where
$\upsilon:X\rightarrow K$ is a Borel map. Then, for any non-negative Borel
function $f$ on $G$, by Tonelli's theorem and the left-invariance of
$\mu_K$, we have:
\begin{eqnarray*}
\int_{X\times K} f(\sigmas(x)\,k)\ d\mu_X\!\otimes\!\mu_K(x,k)
\!\! & = & \!\!
\int_{X\times K} f(\sigmas(x)\,\upsilon(x)\,k)\
d\mu_X\!\otimes\!\mu_K(x,k)
\\
\!\! & = & \!\!
\int_{X\times K} f(\sigmas^\prime(x)\,k)\ d\mu_X\!\otimes\!\mu_K(x,k).
\end{eqnarray*}
This shows that the image measure through $\gamma_\sigmas$ of
$\mu_X\!\otimes\!\mu_K$ does not depend on the choice of $\sigmas$ and
the proof is complete.~$\blacksquare$

From this point onwards, we will assume that the normalization of
the Haar measures $\mu_X$ and $\mu_K$ has been fixed in such a way that
equation~{(\ref{decomposition})} is satisfied.
It will be an easy task, now, to prove the announced result.

\begin{proposition} \label{compact}
If the representation $U$ is square integrable,
every $U$-central subgroup of $G$ is compact.
Hence, in particular,
$G$ admits square integrable irreducible unitary representations
only if $\maxi$ and $K_0$ are compact.
\end{proposition}

{\bf Proof\, :} Indeed, if $U$ is square integrable,
for any $\phi\in\mathcal{H}$ and
$\psi\in\mathcal{A}(U)$, $\phi,\psi\neq 0$, we have:
\[
0<\alpha\equiv\int_G |c_{\psi,\phi}^U(g)|^2\ d\mu_G(g)<+\infty.
\]
Then, for any $U$-central subgroup $K$ of $G$,
using formula~{(\ref{decomposition})}, we find:
\begin{eqnarray*}
0 < \alpha
\!\! & = & \!\!
\int_{X\times K}|c_{\psi,\phi}^U(\sigmas(x)\,k)|^2\ d\mu_X\!\otimes\mu_K(k)
\\
(\mbox{$K$ is $U$-central})
\!\! & = & \!\!
\int_{X\times K} |c_{\psi,\phi}^U(\sigmas(x))|^2\ d\mu_X\!\otimes\!\mu_K
(x,k)
\\
(\mbox{Tonelli's theorem})
\!\! & = & \!\!
\mu_K(K)\,\int_X |c_{\psi,\phi}^U(\sigmas(x))|^2\ d\mu_X(x) < +\infty .
\end{eqnarray*}
It follows that $\mu_K(K)<+\infty$, hence $K$ must be compact.~$\blacksquare$

Our next step will be to show that one can associate, in a natural
way, with the unitary representation $U$ of $G$ a projective
representation of $X$. To this aim, let us define a map
$\emme_\sigmas\, : X\times X\rightarrow \mathbb{T}$ by
\begin{equation}
\emme_\sigmas(x_1,x_2):=\chi(\kappa_\sigmas(x_1,x_2)),
\end{equation}
where we recall that $\chi : K\rightarrow\mathbb{T}$ is the
continuous group homomorphism determined by the restriction of $U$ to
the $U$-central subgroup $K$.

\begin{proposition}
Given a Borel section $\sigmas : X \rightarrow G$, 
the function $\emme_\sigmas$ is a multiplier for $X$ and
the map $P_\sigmas  : X\rightarrow\mathcal{U}(\mathcal{H})$, defined by
\begin{equation}
P_\sigmas(x):=U(\sigmas(x)),\ \ \ \ \forall x\in X,
\end{equation}
is an irreducible projective representation with multiplier $\emme_\sigmas$.
Moreover, if $\sigmas^\prime : X\rightarrow G$ is another Borel section,
$P_\sigmas$ and $P_{\sigmas^\prime}$ are ray equivalent projective
representations, hence,
the multipliers $\emme_\sigmas$ and $\emme_{\sigmas^\prime}$ are similar.
\end{proposition}

{\bf Proof\, :} As $\kappa_\sigmas$ is a Borel map, the function
$\emme_\sigmas$ is Borel and, since
$\kappa_\sigmas(e,x)=\kappa_\sigmas(x,e)=e$,
where here $e$ denotes the identity
both in $X$ and in $G$, we have that $\emme_\sigmas(e,x)=
\emme_\sigmas(x,e)=1$, for any $x\in X$. Besides, since
\begin{eqnarray*}
\sigmas(x_1 x_2 x_3)
\!\! & = & \!\!
\sigmas(x_1)\,\sigmas(x_2)\,\sigmas(x_3)\,
\kappa_\sigmas(x_2,x_3)\,\kappa_\sigmas(x_1,x_2 x_3)
\\
\!\! & = & \!\!
\sigmas(x_1)\,\sigmas(x_2)\,\kappa_\sigmas(x_1,x_2)\,
\sigmas(x_3)\,\kappa_\sigmas(x_1 x_2,x_3),
\end{eqnarray*}
using the fact that the restriction of $U$ to $K$ is the scalar
representation $\chi\,I$, we find:
\begin{eqnarray*}
\chi(\kappa_\sigmas(x_1,x_2 x_3))\, \chi(\kappa_\sigmas(x_2,x_3))\, I
\!\! & = & \!\!
U(\sigmas(x_1x_2x_3))\, U(\sigmas(x_1)\,\sigmas(x_2)\,\sigmas(x_3))^{-1}
\\
\!\! & = & \!\!
\chi(\kappa_\sigmas(x_1x_2,x_3))\chi(\kappa_\sigmas(x_1,x_2))\, I,
\end{eqnarray*}
for all $x_1,x_2,x_3\in X$.
Thus $\emme_\sigmas$ is a multiplier for $X$. Moreover, 
observe that $P_\sigmas$ is a weakly Borel map, $P_\sigmas(e)=I$ and,
setting $x_3=e$ above, we obtain:
\[
U(\sigmas(x_1x_2))=\chi(\kappa_\sigmas(x_1,x_2))\, U(\sigmas(x_1))\,
U(\sigmas(x_2)).
\]
Hence, $P_\sigmas$ is a projective representation
with multiplier $\emme_\sigmas$. The relation
\[
\varpi(P_\sigmas(X))=\varpi(U(G))\subset\mathcal{P}(\mathcal{H})
\]
implies that $P_\sigmas$ is irreducible, since $U$ is.
If $\sigmas^\prime$ is another Borel section,
there is a Borel map $\upsilon : X\rightarrow K$ such that
\[
\sigmas^\prime(x)=\sigmas(x)\,\upsilon(x),\ \ \ \ \forall x\in X.
\]
Then, setting $\beta:=\chi\circ \upsilon$, there exists
a Borel map $\beta : X\rightarrow\mathbb{T}$ such that
\[
P_{\sigmas^\prime}(x)=\beta(x)\, P_\sigmas(x), \ \ \ \ \forall x \in X,
\]
hence, $\varpi(P_{\sigmas^\prime}(x))=\varpi(P_\sigmas(x))$,
for any $x\in X$; namely, $P_\sigmas$ and $P_{\sigmas^\prime}$ are ray
equivalent projective representations.
It follows that the associated multipliers must be similar. In fact,
explicitly, we have:
\begin{eqnarray*}
\chi(\kappa_{\sigmas^\prime}(x_1,x_2))\, I
\!\! & = & \!\!
\chi(\sigmas^\prime(x_1 x_2)\,\sigmas^\prime(x_2)^{-1}
\sigmas^\prime(x_1)^{-1})\, I
\\
\!\! & = & \!\!
\beta(x_1x_2)\,\beta(x_1)^{-1}\beta(x_2)^{-1}\, U(\sigmas(x_1x_2)\,
\sigmas(x_2)^{-1}\sigmas(x_2)^{-1})
\\
\!\! & = & \!\!
\beta(x_1x_2)\,\beta(x_1)^{-1}\beta(x_2)^{-1}
\chi(\kappa_\sigmas(x_1,x_2))\, I,
\end{eqnarray*}
for all $x_1,x_2\in X$.
Thus $\emme_\sigmas$ and $\emme_{\sigmas^\prime}$ are similar multipliers
and the proof is complete.~$\blacksquare$

Since $P_\sigmas$ is a projective representation
with multiplier $\emme_\sigmas$,
one can define, as it has been shown in section~{\ref{3}},
the l.c.s.c.\ group $X_{\emme_\sigmas}$ and the
strongly continuous unitary representation
\begin{equation}
U_{P_\sigmas}  : X_{\emme_\sigmas}\ni (\tau,x)\mapsto \tau^{-1} P_\sigmas(x)\in
\mathcal{U}(\mathcal{H}).
\end{equation}
$U_{P_\sigmas}$ is irreducible since $P_\sigmas$ is. Again, we stress that
this construction does not depend essentially on the choice of the
Borel section $\sigmas$. Indeed, if $\sigmas^\prime$ is another Borel
section, $\emme_\sigmas$ and $\emme_{\sigmas^\prime}$ are similar
multipliers, so that
$X_{\emme_\sigmas}$ and $X_{\emme_{\sigmas^\prime}}$ are
isomorphic topological groups and the representations
$U_{P_\sigmas}$, $U_{P_{\sigmas^\prime}}$ can be identified
under this isomorphism.

At this point, we want to introduce the notion of square integrable
representation modulo a relatively central subgroup.
To this aim, let us recall that, for any compact subset $\mathcal{C}$
of $G$, the set $\mathcal{C}\, K$ is closed (since $K$ is closed) and
the subset $\mathsf{p}(\mathcal{C})$ of $X$ is compact
(since $\mathsf{p}$ is continuous).
We recall also that a Borel measure on a l.c.s.c.\ topological
space is a Radon measure if and only if it
is finite on compact sets (see, for instance, \cite{Gerald},
chapter~7).
Moreover, any Radon measure on a l.c.s.c.\ topological space is regular (in
particular, Haar measures on l.c.s.c.\ groups are regular Radon measures).
Then, let us define $\mathcal{M}_{G,K}$ as the set of the Borel
measures $\mu_{G,K}$ on $G$ that verify the following conditions:
\begin{description}

\item[(a)]\ $\mu_{G,K}$ is absolutely continuous with respect to
the Haar measure $\mu_G$: $\mu_{G,K}\ll \mu_G$ (i.e.\ $\mu_{G,K}(
\mathcal{B})=0$ for every Borel set $\mathcal{B}\subset G$ such that
$\mu_G(\mathcal{B})=0$); 

\item[(b)]\ for any compact subset $\mathcal{C}$ of $G$,
\begin{equation} \label{fix}
\mu_{G,K}(\mathcal{C}\, K)= \mu_X(\mathsf{p}(\mathcal{C})).
\end{equation}
\end{description}

Notice that, since for any compact subset $\mathcal{C}$ of $G$
the subset $\mathsf{p}(\mathcal{C})$ of $X$ is compact, hence
\[
\mu_{G,K}(\mathcal{C})\le\mu_{G,K}(\mathcal{C}\, K)<+\infty,\ \ \ \
\forall \mu_{G,K}\in\mathcal{M}_{G,K},
\]
the set $\mathcal{M}_{G,K}$ consists of (regular) Radon measures.
Moreover, any measure $\mu_{G,K}$ in $\mathcal{M}_{G,K}$ is nonzero.
Indeed, given a compact subset $\mathcal{X}$ of $X$, it is a standard
result (see, for instance,~\cite{Raja},
chapter~V, sect.~4) that there is a compact subset
$\mathcal{C}$ of $G$ such that $\mathcal{X}=\mathsf{p}(\mathcal{C})$.
Then, choose the compact subset $\mathcal{X}$ so that
$\mu_X(\mathcal{X})>0$ (for example, take the closure of a 
nonempty precompact open set); hence: $\mu_{G,K}(\mathcal{C}\, K)>0$. \\
Clearly, for any couple of positive constants $\alpha_1,\alpha_2$, with
$\alpha_1+\alpha_2=1$, and any couple of measures
$\mu_{G,K}^{1},\,\mu_{G,K}^{2}\in\mathcal{M}_{G,K}$, the measure
$\alpha_1\,\mu_{G,K}^{1}+\alpha_2\,\mu_{G,K}^{2}$ belongs to
$\mathcal{M}_{G,K}$.
Another fundamental property of the set $\mathcal{M}_{G,K}$ is that it is
left-invariant in the following sense. For any
$\mu_{G,K}\in\mathcal{M}_{G,K}$ and any $g\in G$, one can define
the $g$-translate measure $\mu_{G,K}^g$ by
\begin{equation}
\mu_{G,K}^g(\mathcal{B})=\mu_{G,K}(g\,\mathcal{B}),
\end{equation}
where $\mathcal{B}$ is an arbitrary Borel subset of $G$. Then,
if $\mu_G(\mathcal{B})=0$, we have that $\mu_G(g\,\mathcal{B})=0$ and,
since $\mu_{G,K}\ll\mu_G$,
\[
\mu_{G,K}^g(\mathcal{B})=\mu_{G,K}(g\,\mathcal{B})=0;
\]
hence: $\mu_{G,K}^g\ll\mu_G$.
Besides,
for any compact subset $\mathcal{C}$ of $G$, the set $g\,\mathcal{C}$
is compact and
\[
\mu_{G,K}^g(\mathcal{C}\, K)=\mu_{G,K}(g\,\mathcal{C}\,K)=
\mu_X(\mathsf{p}(g)\,\mathsf{p}(\mathcal{C}))=
\mu_X(\mathsf{p}(\mathcal{C})).
\]
Thus, the Borel measure $\mu_{G,K}^g$ belongs to $\mathcal{M}_{G,K}$.\\
Equation~{(\ref{fix})} fixes, in particular,
the normalization of the measures in $\mathcal{M}_{G,K}$ in the sense
that, given $\mu_{G,K}\in\mathcal{M}_{G,K}$, the measure
$\alpha\,\mu_{G,K}$, with $0<\alpha\neq 1$, does not
belong to $\mathcal{M}_{G,K}$.
We could have defined $\mathcal{M}_{G,K}$ letting this normalization free,
but this would have introduced cumbersome constants in many
formulae. Let us now give a complete cheracterization of the set
$\mathcal{M}_{G,K}$.

\begin{proposition} \label{measure}
The left-invariant set of nonzero Radon measures
$\mathcal{M}_{G,K}$ is not empty. Any measure $\mu_{G,K}$ in
$\mathcal{M}_{G,K}$ is of the form $d\mu_{G,K}=\varrho\,d\mu_G$, where
$\varrho:G\rightarrow\mathbb{R}$ is a non-negative Borel function ---
which is essentially unique, i.e.\ unique modulo alterations on
$\mu_G$-null sets ---
verifying the following property: 
\begin{equation} \label{density}
\int_K \varrho(\sigmas(x)\, k)\ d\mu_K(k)=1,\ \ \
\mbox{for $\mu_X$-almost all}\ x\in X;
\end{equation}
here the integral does not depend on the choice of the Borel
section $\sigmas$. Conversely, every
Borel measure $\mu_{G,K}$ on $G$ such that
$d\mu_{G,K}=\varrho\,d\mu_G$, for some non-negative Borel function
$\varrho:G\rightarrow\mathbb{R}$ 
verifying property~{(\ref{density})}, is contained in $\mathcal{M}_{G,K}$.
\end{proposition}

{\bf Proof\, :} Let $\mu_{G,K}$ be a measure in $\mathcal{M}_{G,K}$.
Then, since $\mu_{G,K}\ll \mu_G$, according to the Radon-Nikodym
theorem, $d\mu_{G,K}=\varrho\, d\mu_G$, for some essentially unique
non-negative Borel function $\varrho$ on $G$.
Now, take a compact subset $\mathcal{C}$ of $G$ and denote by
$\mathcal{X}$ the compact subset $\mathsf{p}(\mathcal{C})$
of $X$ (we have already recalled that each compact subset of $X$ can be
obtained in this way).
Besides, for any set $\mathcal{Y}$, denote by
$\imath_\mathcal{Y}$ its indicator function.
At this point, we have:
\begin{eqnarray*}
\mu_{G,K}(\mathcal{C}\,K)
\!\! & = & \!\!
\int_G \imath_{\mathcal{C}K}(g)\ d\mu_{G,K}(g)
\\
\!\! & = & \!\!
\int_G \imath_{\mathcal{C}K}(g)\, \varrho(g)\ d\mu_G(g)
\\
\!\! & = & \!\!
\int_{X\times K}\imath_{\mathcal{C}K}(\sigmas(x)\,k)\,
\varrho(\sigmas(x)\,k)\ d\mu_X\!\otimes\!\mu_K(x,k),
\end{eqnarray*}
where for obtaining the last equality we have used
formula~{(\ref{decomposition})}. Then, since
\[
\imath_{\mathcal{C}K}(\sigmas(x)\,k)=
\imath_{\mathcal{X}\times K}(x,k)=\imath_\mathcal{X}(x),\ \ \ \
\forall(x,k)\in X\times K,
\]
by virtue of Tonelli's theorem, we find:
\[
\mu_{G,K}(\mathcal{C}\,K)= \int_{\mathcal{X}}
\bar{\varrho}(x)\ d\mu_X(x),
\]
where $\bar{\varrho}$ is the Borel function defined by
\[
\bar{\varrho}(x):=\int_K \varrho(\sigmas(x)\, k)\ d\mu_K(k).
\]
Here we stress that, by the left-invariance of $\mu_K$, the Borel function
$\bar{\varrho}$ does not depend on the choice of $\sigmas$.
Hence, for any compact subset $\mathcal{C}$ of $G$, we have:
\[
\mu_X(\mathcal{X})=\mu_{G,K}(\mathcal{C}\,K)=
\int_{\mathcal{X}} \bar{\varrho}(x)\
d\mu_X(x).
\]
Then, since $\mu_X$ is a regular measure (in particular, inner regular),
it is determined uniquely by
its value on compact sets and, by the arbitariness of
the compact set $\mathcal{X}$,
we argue that $\bar{\varrho}(x)=1$, for $\mu_X$-almost $x$ in $X$.\\
Conversely, reasoning as above, one finds that
every non-negative Borel function $\varrho$ on $G$ satisfying
property~{(\ref{density})} defines a measure $\mu_{G,K}$
of the form $d\mu_{G,K}=\varrho\,d\mu_G$ which belongs to
$\mathcal{M}_{G,K}$. In particular, this is true
for every continuous function $\varrho:G\rightarrow\mathbb{R}$ such that
\begin{equation} \label{strong}
\int_K \varrho(g\, k)\ d\mu_K(k)=1,\ \ \ \ \forall g\in G.
\end{equation}
Now, such a function does exist (see, for instance, Proposition~2,
p.~258, of ref.~\cite{Gaal}), so that $\mathcal{M}_{G,K}$ is a
nonempty set.~$\blacksquare$

\begin{remark}
Condition~{(\ref{density})} is equivalent to the following one:
\begin{equation} \label{density2}
\int_K\varrho(g\,k)\ d\mu_K(k)=1,\ \ \
\mbox{for $\mu_G$-almost all}\ g\in G.
\end{equation}
In fact, if we define the sets
\begin{eqnarray*}
\noG
\!\! & := & \!\!
\Big\{g\in G\,|\ \int_K\varrho(g\,k)\ d\mu_K(k)\neq 1\Big\},
\\
\noX
\!\! & := & \!\!
\Big\{x\in X\,|\ \int_K\varrho(\sigmas(x)\,k)\ d\mu_K(k)\neq 1\Big\},
\end{eqnarray*}
by the left-invariance of $\mu_K$, we have that $\noX=\mathsf{p}(\noG)$ and
$\noG=\noG\, K=\mathsf{p}^{-1}(\noX)$, hence $\noG=\gamma_\sigmas(\noX,K)$.
Thus, by Lemma~{\ref{main}},
\[
\mu_G(\noG)=0\ \ \Longleftrightarrow \ \ \mu_X(\noX)=0.
\]
\end{remark}

We will call the essentially unique non-negative Borel function $\varrho$ of
Proposition~{\ref{measure}} the {\it density canonically
associated with} $\mu_{G,K}\in\mathcal{M}_{G,K}$.
Notice that, for every measure $\mu_{G,K}$ in $\mathcal{M}_{G,K}$ and any
$g\in G$, the $g$-translate measure $\mu_{G,K}^g$ is of the form
$d\mu_{G,K}^g = \varrho^g\,d\mu_G$, where $\varrho^g$ is the
$g$-translate of density $\varrho$ canonically associated with $\mu_{G,K}$,
i.e.\ the Borel function defined by $\varrho^g(g^\prime)=\varrho(gg^\prime)$.
In fact, for any Borel set $\mathcal{B}\subset G$, denoted by
$\imath_\mathcal{B}$ the indicator function of $\mathcal{B}$, we have:
\begin{eqnarray*}
\mu_{G,K}^g(\mathcal{B})
\!\! & = & \!\!
\int_G \imath_{g\mathcal{B}}(g^\prime)\,\varrho(g^\prime)\ d\mu_G(g^\prime)
\\
\!\! & = & \!\!
\int_G \imath_{\mathcal{B}}(g^{-1}g^\prime)\,
\varrho(g^\prime)\ d\mu_G(g^\prime)
\\
\!\! & = & \!\!
\int_G \imath_{\mathcal{B}}(g^\prime)\,\varrho(gg^\prime)\ d\mu_G(g^\prime).
\end{eqnarray*}
Then, consider the set $\mathfrak{F}_{G,K}$ of classes of Borel functions
on $G$ defined as follows. Each element of $\mathfrak{F}_{G,K}$
consists of all real-valued Borel functions on $G$
that are $\mu_G$-almost everywhere equal to 
a non-negative Borel function $\varrho$
satisfying condition~{(\ref{density2})}; namely,
this element is the equivalence class of $\varrho$, in the linear space of
real-valued Borel functions on $G$,
with respect to $\mu_G$. For any $g\in G$,
one can define the $g$-translate of an element of $\mathfrak{F}_{G,K}$
as the equivalence class, with respect to $\mu_G$, of the
$g$-translate of a representative function of this element. As we have just
seen, this equivalence class is again an element of $\mathfrak{F}_{G,K}$.
In this sense, the set $\mathfrak{F}_{G,K}$ is left-invariant
and, according to Proposition~\ref{measure}, there is
a one-to-one correspondence between classes of functions in
$\mathfrak{F}_{G,K}$ and measures in $\mathcal{M}_{G,K}$.
From the proof of Proposition~\ref{measure}, it turns out that
there are elements of $\mathfrak{F}_{G,K}$ that admit a
continuous representative function satisfying the stronger
condition~{(\ref{strong})}. We will call {\it regular}
the corresponding measures in $\mathcal{M}_{G,K}$. It is clear
that regular measures form a left-invariant subset of $\mathcal{M}_{G,K}$.

Now, let us define the following subset of the Hilbert space $\mathcal{H}$
(namely, the set of `admissible vectors for $U$ modulo $K$'):
\[
\mathcal{A}(U,K)
:=\left\{\psi\in\mathcal{H}\,|\ \exists\ \phi\in\mathcal{H},\,
\mu_{G,K}\in\mathcal{M}_{G,K}:\, \phi\neq 0,\,
c_{\psi,\phi}^{U} \in L^2(G,\mu_{G,K})\right\},
\]
which will be shown to be a linear manifold.
We will say that the representation $U$ of $G$ is
{\it square integrable modulo the relatively central subgroup} $K$ if
\[
\mathcal{A}(U,K)\neq\{0\}.
\]
Eventually, we are ready to establish the central results of this section.

\begin{proposition} \label{equivalence}
The strongly continuous irreducible unitary representation
$U$ of the l.c.s.c.\ group $G$ is square integrable modulo
the relatively central subgroup $K$ if and
only if $P_\sigmas$ is a square integrable projective representation of $X$,
hence, if and only if $U_{P_\sigmas}$ is a square integrable
unitary representation of $X_{\emme_\sigmas}$. Moreover, the respective
sets of admissible vectors coincide:
\[
\mathcal{A}(U,K)=\mathcal{A}(P_\sigmas)=\mathcal{A}(U_{P_\sigmas}).
\]
\end{proposition}

{\bf Proof\, :} Let $\mu_{G,K}$ be a measure in $\mathcal{M}_{G,K}$.
Then, denoted by $\varrho$ the density canonically associated with
$\mu_{G,K}$, we have:
\begin{eqnarray*}
\int_G |c_{\psi,\phi}^U(g)|^2\ d\mu_{G,K}(g)
\!\! & = & \!\!
\int_{X\times K}\!\! |\langle U(\sigmas(x)\,k)\,\psi,\phi\rangle|^2\,
\varrho(\sigmas(x)\,k)\
d\mu_X\!\otimes\!\mu_K(x,k)
\\
\!\! & = & \!\!
\int_{X\times K}\! |c_{\psi,\phi}^U(\sigmas(x))|^2\,\varrho(\sigmas(x)\,k)\
d\mu_X\!\otimes\!\mu_K(x,k)
\\
\!\! & = & \!\!
\int_X |c_{\psi,\phi}^{P_\sigmas}(x)|^2\ d\mu_X(x),\ \ \ \
\forall\, \psi,\phi\in\mathcal{H}.
\end{eqnarray*}
It follows that $\mathcal{A}(U,K)=\mathcal{A}(P_\sigmas)=
\mathcal{A}(U_{P_\sigmas})$, hence $U$ is square integrable modulo $K$
if and only if $P_\sigmas$ (or $U_{P_\sigmas}$) is square
integrable.~$\blacksquare$

We can now prove a generalization of the theorem of Duflo and Moore.

\begin{theorem} \label{Duflo3}
Let the representation $U$ of $G$ be square integrable modulo
the relatively central subgroup $K$. 
Then, for any $\phi\in\mathcal{H}$
and any $\psi$ in the dense linear manifold $\mathcal{A}(U,K)$, the
cofficient $c_{\psi,\phi}^{U}$ is square integrable with respect to
any measure $\mu_{G,K}$ in $\mathcal{M}_{G,K}$. For any
nonzero vector $\psi\in\mathcal{A}(U,K)$, the map
\begin{equation}
C_{\psi}^{U,K} :\ \mathcal{H}\ni\phi\mapsto c_{\psi,\phi}^{U}
\in L^{2}(G,\mu_{G,K})
\end{equation}
defines a linear operator which is essentially isometric.
Furthermore, there exists a unique positive selfadjoint injective linear
operator $D_{U,K}$ in $\mathcal{H}$ such that
\[
\mathcal{A}(U,K)=\mathrm{Dom}\left(D_{U,K}\right) 
\]
and
\begin{equation} 
\int_X c_{\psi_1,\phi_1}^{U}(g)^{*}\,
c_{\psi_2,\phi_2}^{U}(g)\ d\mu_{G,K}(g) = 
\langle\phi_1 ,\phi_2\rangle\,
\langle D_{U,K}\,\psi_2, D_{U,K}\,\psi_1\rangle ,
\end{equation}
for all $\phi_1,\phi_2\in\mathcal{H}$, for all $\psi_1,\psi_2\in
\mathcal{A}(U,K)$. Finally, the operator $D_{U,K}$ does not depend
on the choice of the measure $\mu_{G,K}$ in $\mathcal{M}_{G,K}$
and, if $D_{P_\sigmas}$ is normalized according to $\mu_X$,
$D_{U,K}=D_{P_\sigmas}$.
Thus, $D_{U,K}$ is bounded if and only if
$X=G/K$ is unimodular and, in such case, it is a multiple of the identity.
\end{theorem}

{\bf Proof\, :} If $U$ is square integrable modulo $K$, then,
for any $\phi\in\mathcal{H}$, $\psi\in\mathcal{A}(U,K)=
\mathcal{A}(P_\sigmas)$ and $\mu_{G,K}\in\mathcal{M}_{G,K}$,
we have:
\[
\int_G |c_{\psi,\phi}^U(g)|^2\ d\mu_{G,K}(g)=
\int_X |c_{\psi,\phi}^{P_\sigmas}(x)|^2\ d\mu_X(x)<+\infty.
\]
It follows that
\begin{eqnarray*}
+\infty
\!\! & > & \!\!
\int_G \langle\phi_1,U(g)\,\psi_1\rangle \langle U(g)\,\psi_2,\phi_2\rangle\
d\mu_{G,K}(g)
\\
\!\! & = & \!\!
\int_{X\times K}\!\!\! \langle\phi_1,P_\sigmas(x)\,\psi_1\rangle
\langle P_\sigmas(x)\,\psi_2,\phi_2\rangle\, \varrho(\sigmas(x)\,k)\
d\mu_X\!\otimes\!\mu_K(x,k)
\\
(\mbox{Fubini's th.})
\!\! & = & \!\!
\int_X c_{\psi_1,\phi_1}^{P_\sigmas}(x)^*\, c_{\psi_2,\phi_2}^{P_\sigmas}
(x) \left(\int_K \varrho(\sigmas(x)\,k)\ d\mu_K(k)\right) d\mu_X(x)
\\
\!\! & = & \!\!
\int_X c_{\psi_1,\phi_1}^{P_\sigmas}(x)^*\, c_{\psi_2,\phi_2}^{P_\sigmas}
(x)\ d\mu_X(x)
\\
(\mbox{Theorem~{\ref{Duflo2}}})
\!\! & = & \!\!
\langle\phi_1,\phi_2\rangle\, \langle D_{P_\sigmas}\,\psi_2,
D_{P_\sigmas}\,\psi_1\rangle,
\end{eqnarray*}
$\forall\, \phi_1,\phi_2\in\mathcal{H},\ \forall\, \psi_1,\psi_2\in
\mathcal{A}(P_\sigmas)=\mathcal{A}(U,K)$. 
Now, set $D_{U,K}= D_{P_\sigmas}$ and notice that nothing in the
preceding arguments depends on the choice of $\mu_{G,K}$ in
$\mathcal{M}_{G,K}$. This completes the proof.~$\blacksquare$

\begin{remark} \label{measures}
Assume that $K$ is compact. Then, if we set as usual $\mu_K(K)=1$,
$\mu_G$ belongs to $\mathcal{M}_{G,K}$.
Indeed, denoted by $\imath_\mathcal{Y}$ the indicator function of a set
$\mathcal{Y}$, for any compact subset $\mathcal{C}$ of $G$, we have:
\[
\imath_{\mathcal{C}K}(\sigmas(x)\,k)=\imath_{\mathsf{p}(\mathcal{C})\times
K}(x,k)=\imath_{\mathsf{p}(\mathcal{C})}(x),\ \ \ \
\forall x\in X,\ \forall k\in K.
\]
Hence:
\[
\mu_G(\mathcal{C}\, K)=\int_{X\times K}\imath_{\mathcal{C}K}
(\sigmas(x)\,k)\ d\mu_X\!\otimes\mu_K(x,k)=\mu_X(\mathsf{p}(\mathcal{C}))\,
\mu_K(K).
\]
The same argument shows that, if $\mu_G$ belongs to $\mathcal{M}_{G,K}$,
then $\mu_K(K)=1$ and $K$ must be compact.
\end{remark}

\begin{remark} \label{comp}
From Remark~\ref{measures} it follows that, if $K$ is a compact $U$-central
subgroup, $U$ is square integrable if and only if it is
square integrable modulo $K$.
In particular, the square-integrability of a unitary
representation is equivalent to the square-integrability
modulo the trivial subgroup $\{e\}$.
\end{remark}

\begin{remark}
If $K$ is a closed central subgroup of $G$, we can prove that
the quotient group $X$ is
unimodular if and only if $G$ is. Indeed, if $K$ is central, then
$\Delta_G(k)=\Delta_K(k)=1$, for all $k\in K$, and,
for any $f\in L^1(G,\mu_G)$, using Fubini's theorem we have:
\begin{eqnarray*}
\int_G f(g)\ d\mu_G(g)
\!\! & = & \!\!
\int_{X\times K}\!\!\! f(\sigmas(x)\, k)\ d\mu_X\!\otimes\!\mu_K(x,k)
\\
(\mbox{$K$ {\rm unimodular}})
\!\! & = & \!\!
\int_{X\times K}\!\!\!\! f(\sigmas(x^{-1})\, k^{-1})\,
\Delta_X(x^{-1})\ d\mu_X\!\otimes\!\mu_K(x,k)
\\
(\mbox{$K$ {\rm central}})
\!\! & = & \!\!
\int_{X\times K}\!\!\!\! f(k^{-1}\sigmas(x^{-1}))\,\Delta_X(x^{-1})\
d\mu_X\!\otimes\!\mu_K(x,k)
\\
(\sigmas(x^{-1})=\kappa_\sigmas(x^{-1},x)^{-1}\sigmas(x)^{-1})
\!\! & = & \!\!
\int_{X\times K}\!\!\! f(k^{-1}\sigmas(x)^{-1})\,\Delta_X(x^{-1})\
d\mu_X\!\otimes\!\mu_K(x,k)
\\
(\mathsf{p}(g^{-1})=\mathsf{p}(g)^{-1})
\!\! & = & \!\!
\int_G f(g^{-1})\,\Delta_X(\mathsf{p}(g^{-1}))\ d\mu_G (g),
\end{eqnarray*}
where $\Delta_X$ is the modular function on $X$.
Hence:
\begin{equation}
\Delta_G(g)=\Delta_G(g\,k)=
\Delta_X(\mathsf{p}(g)),\ \ \ \ \forall g\in G,\ \forall k\in K,
\end{equation}
so that $X$ is unimodular if and only if $G$ is.
\end{remark}

We will now prove some properties concerning the square-integrability
of a unitary representation modulo a relatively
central subgroup.

\begin{proposition}
If $U$ is a square integrable representation of $G$, then $U$ is
square integrable modulo every $U$-central subgroup of $G$.
\end{proposition}

{\bf Proof\, :} If $U$ is square integrable, then, according to
Proposition~\ref{compact}, any $U$-central subgroup is compact.
Hence, recalling Remark~\ref{comp}, we conclude that $U$ is square
integrable modulo every $U$-central subgroup.~$\blacksquare$

\begin{proposition}
If $G$ admits a representation $U$ that is square integrable modulo a
compact relatively central subgroup, then every $U$-central
subgroup of $G$ is compact.
\end{proposition}

{\bf Proof\, :} If $U$ is square integrable modulo a compact
relatively central subgroup, then, according to Remark~\ref{comp},
it is square integrable, hence, by Proposition~\ref{compact},
every $U$-central subgroup is compact.~$\blacksquare$

The following result can be regarded as a generalization of
Remark~{\ref{measures}}.

\begin{proposition}
Let $K,\tilde{K}$ be $U$-central subgroups of $G$, with $\tilde{K}\subset K$.
Denote by $\tilde{X},\breve{X}$ respectively the quotient groups
$G/\tilde{K}$ and $K/\tilde{K}$. Then, if $\breve{X}$ is compact, for a
suitable normalization of the left Haar measures $\mu_{\tilde{X}}$ and
$\mu_{\tilde{K}}$, we have that
\begin{equation} \label{sub}
\mathcal{M}_{G,\tilde{K}}\subset
\mathcal{M}_{G,K}.
\end{equation}
Hence, if $\breve{X}$ is compact, $U$ is square integrable modulo
$\tilde{K}$ if and only if it is square integrable modulo $K$.
Conversely, if relation~{(\ref{sub})} holds, then $\breve{X}$
must be compact.
\end{proposition}

{\bf Proof\, :}
Suppose that $\breve{X}$ is compact.
Then we can set the normalization of the Haar
measure on it as usual: $\mu_{\breve{X}}(\breve{X})=1$. This choice
fixes the normalization of the Haar measures
$\mu_{\tilde{K}}$ and $\mu_{\tilde{X}}$, if, denoted by
$\breve{\sigmas}:\breve{X}\rightarrow K$ and $\tilde{\sigmas}:\tilde{X}
\rightarrow G$ a couple of Borel sections, one imposes that the image
measure, through the Borel isomorphism
$\gamma_{\sigmab}:\breve{X}\times\tilde{K}\rightarrow K$, of
$\mu_{\breve{X}}\!\otimes\!\mu_{\tilde{K}}$ is $\mu_K$ and
$\mu_G$ is the image measure of $\mu_{\tilde{X}}\!\otimes\!\mu_{\tilde{K}}$
through $\gamma_{\tilde{\mathsf{s}}}:\tilde{X}\times\tilde{K}\rightarrow G$.
Now, let $\tilde{\mu}_{G,\tilde{K}}$
be a regular measure in $\mathcal{M}_{G,\tilde{K}}$ and 
$\tilde{\varrho}$ the density canonically associated with it, which can be
assumed to be a non-negative continuous function such that
\begin{equation} \label{regular}
\int_{\tilde{K}} \tilde{\varrho}(g\,\tilde{k})\ d\mu_{\tilde{K}}
(\tilde{k})=1,\ \ \ \ \forall g\in G.
\end{equation}
Then, for any compact
subset $\mathcal{C}$ of $G$, denoted by $\imath_{\mathcal{C}K}$ the
indicator function of the set $\mathcal{C}\,K$, we have:
\begin{eqnarray*}
\tilde{\mu}_{G,\tilde{K}}(\mathcal{C}\,K)
\!\! & = & \!\!
\int_G \imath_{\mathcal{C} K}(g)\,\tilde{\varrho}(g)\
d\mu_G(g)
\\
\!\! & = & \!\!
\int_{X\times K} \imath_{\mathcal{C}K}(\sigmas(x)\,k)\,\tilde{\varrho}
(\sigmas(x)\,k)\ d\mu_{X}\!\otimes\!\mu_{K}(x,k)
\\
\!\! & = & \!\!
\int_{X\times K} \imath_{\mathsf{p}(\mathcal{C})}(x)\,\tilde{\varrho}
(\sigmas(x)\,k)\ d\mu_X\!\otimes\!\mu_K(x,k).
\end{eqnarray*}
Next, by Lemma~{\ref{main}} and Tonelli's theorem,
the integral on $X\times K$ can be further decomposed and we get:
\begin{eqnarray*}
\tilde{\mu}_{G,\tilde{K}}(\mathcal{C}\, K)
\!\! & = & \!\!
\int_{X\times\breve{X}\times\tilde{K}} \imath_{\mathsf{p}(\mathcal{C})}
(x)\,
\tilde{\varrho}(\sigmas(x)\,\sigmab(\breve{x})\,
\tilde{k})\ d\mu_{X}\!\otimes\!\mu_{\breve{X}}\!\otimes\!\mu_{\tilde{K}}
(x,\breve{x},\tilde{k})
\\
(\mbox{property~{(\ref{regular})}})
\!\! & = & \!\!
\int_{X\times\breve{X}} \imath_{\mathsf{p}(\mathcal{C})}(x)
\ d\mu_{X}\!\otimes\!\mu_{\breve{X}}(x,\breve{x}).
\end{eqnarray*}
It follows that $\tilde{\mu}_{G,\tilde{K}}(\mathcal{C}\, K)
=\mu_X(\mathsf{p}(\mathcal{C}))$,
for any compact subset $\mathcal{C}$ of $G$, hence,
$\tilde{\mu}_{G,\tilde{K}}$ belongs to $\mathcal{M}_{G,K}$.\\
In passing, notice that, by the same argument, if
$\mathcal{M}_{G,\tilde{K}}$ is a subset of $\mathcal{M}_{G,K}$, then,
for a suitable normalization of the left Haar measures
$\mu_{\tilde{X}}$ and $\mu_{\tilde{K}}$, it turns out that
$\mu_{\breve{X}}(\breve{X})=1$, thus $\breve{X}$ must be compact. \\
Now, let $\mu_{G,\tilde{K}}$ be any measure in $\mathcal{M}_{G,\tilde{K}}$.
Then, observing that $\mathcal{C}\,K=\mathcal{C}\,K \tilde{K}$,
since $\tilde{K}\subset K$, denoted by $\varrho$ the density
canonically associated with $\mu_{G,\tilde{K}}$
and by $\tilde{\mathsf{p}}$ the projection homomorphism
of $G$ onto $\tilde{X}$, we have:
\begin{eqnarray*}
\mu_{G,\tilde{K}}(\mathcal{C}\,K)
\!\! & = & \!\!
\int_{\tilde{X}\times\tilde{K}}\imath_{\mathcal{C}K\tilde{K}}
(\tilde{\sigmas}(\tilde{x})\,\tilde{k})\,\varrho(\tilde{\sigmas}(\tilde{x})\,
\tilde{k})\ d\mu_{\tilde{X}}\!\otimes\!\mu_{\tilde{K}}(\tilde{x},\tilde{k})
\\
\!\!  & = & \!\!
\int_{\tilde{X}\times\tilde{K}}\imath_{\tilde{\mathsf{p}}(\mathcal{C}K)}
(\tilde{x})\,\varrho(\tilde{\sigmas}(\tilde{x})\,\tilde{k})\
d\mu_{\tilde{X}}\!\otimes\mu_{\tilde{K}}(\tilde{x},\tilde{k}).
\end{eqnarray*}
Hence, for any compact subset $\mathcal{C}$ of $G$, 
\[
\mu_{G,\tilde{K}}(\mathcal{C}\,K)=
\mu_{\tilde{X}}(\tilde{\mathsf{p}}(\mathcal{C}\,K))=
\tilde{\mu}_{G,\tilde{K}}(\mathcal{C}\,K)=
\mu_X(\mathsf{p}(\mathcal{C})).
\]
Thus $\mathcal{M}_{G,\tilde{K}}\subset\mathcal{M}_{G,K}$ and
the proof is complete.~$\blacksquare$


\section{Intertwining properties} \label{5}

In this section, we want to investigate the intertwining properties
of the operators $C_\psi^{P_\sigmas}$ and $C_\psi^{U,K}$.
To this aim, let us recall first that with any Borel section
$\sigmas:X\rightarrow G$ one can associate 
the map $\mathsf{c}_\sigmas\,: G\times X\rightarrow K$, defined by
\begin{equation}
\mathsf{c}_\sigmas(g,x):= \sigmas(x)^{-1} g^{-1} \, \sigmas(g[x]),
\end{equation}
which is called the cocycle associated with $\sigmas$. 
Recall also (see, for instance, \cite{Raja})
that the representation of $G$ unitarily induced by the representation
$\chi$ of $K$ in $\mathbb{C}$
can be realized as the representation $R^{\chi,\sigmas} :
X\rightarrow \mathcal{U}(L^2(X,\mu_X))$ defined by:
\begin{equation} \label{cond}
\left(R^{\chi,\sigmas}_g\, f\right)(x):=
\chi(\mathsf{c}_\sigmas(g^{-1},x))\, f(g^{-1}[x]),\ \ \ \
f\in L^2(X,\mu_X),
\end{equation}
where we have used the fact that $\mu_X$ is an invariant measure
with respect to the natural action of $G$ on $X$.

Now, let us define $\hilb_0$ as the linear space 
of complex-valued Borel functions
$f$ on $G$ that satisfy the following conditions:
\begin{equation} \label{condition}
f(gk)=\chi(k)^{-1} f(g), \ \ \ \ \forall g\in G,\ \forall k\in K,
\end{equation}
and
\begin{equation} \label{condition*}
\int_X|f(\sigmas(x)|^2\ d\mu_X(x) < +\infty.
\end{equation}
Observe that, due to the first condition, the second one does not depend on
the choice of the Borel section $\sigmas$. Next, identifying two
functions $f,\tilde{f}$ in $\hilb_0$ if
$f(\sigmas(x))=\tilde{f}(\sigmas(x))$ for $\mu_X$-almost all $x\in X$,
we can define the pre-Hilbert space $\hilb$ of equivalence classes
of functions in $\hilb_0$ with scalar product
\begin{equation}
\langle f_1,f_2\rangle_{\hilb} :=
\int_X f_1(\sigmas(x))^*\, f_2(\sigmas(x))\ d\mu_X(x).
\end{equation}
At this point, given a measure $\mu_{G,K}$ in $\mathcal{M}_{G,K}$ and
denoted by $\varrho$ the density canonically associated with $\mu_{G,K}$,
for any $f\in\hilb$, we have:
\begin{eqnarray*}
\|f\|_{\hilb}^2
\!\! & := & \!\!
\int_X|f(\sigmas(x))|^2\ d\mu_X(x)
\\
(\mbox{condition~{(\ref{condition})}, property~{(\ref{density})}})
\!\! & = & \!\!
\int_{X\times K} \!\!\!\! |f(\sigmas(x)\,k)|^2\,\varrho(\sigmas(x)\,k)\
d\mu_X\!\otimes\!\mu_K(x,k).
\end{eqnarray*}
Hence, we find that
\begin{equation} \label{norm}
\|f\|_{\hilb}^2 =
\int_G |f(g)|^2\ d\mu_{G,K}(g),\ \ \ \ \forall f \in\hilb.
\end{equation}
Then, one can associate in a natural way with any function $f$
(representative of an equivalence class of functions) in $\hilb$
its equivalence class of functions
with respect to $\mu_{G,K}$ (i.e.\ the class of all Borel functions
$\mu_{G,K}$-almost everywhere equal to $f$),
so obtaining a linear submanifold
$\hilbm$ of $L^2(G,\mu_{G,K})$ and an isomorphism of pre-Hilbert spaces
$\isom:\hilb\rightarrow\hilbm$. From this point onwards, with the
customary abuse, we will not make a distinction between
an element of $\hilb$ or $\hilbm$ and a representative function of
this element, whenever this distinction will be irrelevant;
thus, in particular, $\isom^{-1}$ will be regarded as the map which
associates with any function in $\hilbm$ a function
$\mu_{G,K}$-a.e.\ equal to it that verifies condition~{(\ref{condition})}.\\
Let us prove that $\hilb$, $\hilbm$
are actually Hilbert spaces and, hence, $\isom$ is a unitary operator.

\begin{proposition} \label{isometry}
For any Borel section $\sigmas :X\rightarrow G$, there is an isometric
linear operator $F_\sigmas : L^2(X,\mu_X)\rightarrow L^2(G,\mu_{G,K})$,
such that $\mathrm{Ran}(F_\sigmas)=\hilbm$, which is explicitly defined by
\begin{equation}
\left(F_\sigmas\, \varphi\right)(g)=\chi(\sigmas(\mathsf{p}(g))^{-1}g)^{-1}\,
\varphi(\mathsf{p}(g)),\ \ \ \ \forall g\in G,
\end{equation}
for any $\varphi\in L^2(X,\mu_X)$. Thus $\hilbm$ is a closed subspace of
$L^2(G,\mu_{G,K})$, hence a separable complex Hilbert space.
\end{proposition}

{\bf Proof\, :} For any $\varphi\in L^2(X,\mu_X)$, since $\gamma_\sigmas$
is a Borel isomorphism, we can define a Borel function $f$ on $G$ setting
\[
f(\gamma_\sigmas(x,k))=\chi(k)^{-1} \varphi(x),
\ \ \ \ \forall x\in X,\ \forall k\in K.
\]
Now, we have:
\begin{eqnarray*}
\int_G |f(g)|^2\ d\mu_{G,K}(g)
\!\! & = & \!\!
\int_{X\times K} |\chi(k)^{-1}\varphi(x)|^2\, \varrho(\sigmas(x)\,k)\
d\mu_X\!\otimes\!\mu_K(x,k)
\\
\!\! & = & \!\!
\int_X |\varphi(x)|^2\ d\mu_X(x).
\end{eqnarray*}
Thus the mapping $\varphi\mapsto f$ defines a linear isometry
$F_\sigmas^0$ from
$L^2(X,\mu_X)$ into $L^2(G,\mu_{G,K})$.
Observe also that
\[
f(\sigmas(x)\,k\,k^\prime)=\chi(k\,k^\prime)^{-1}\varphi(x)=
\chi(k^\prime)^{-1} f(\sigmas(x)k),\ \ \ \
\forall x\in X,\ \forall\, k,k^\prime\in K.
\]
It follows that range of this mapping is contained in $\hilbm$;
let us show that it coincides with $\hilbm$. Indeed, take any
function $f$ in $\hilb$ and set:
\[
\varphi(x)=f(\sigmas(x)), \ \ \ \ \forall x\in X.
\]
Then, $\varphi:X\rightarrow\mathbb{C}$ is a Borel function and
\[
\int_X |\varphi(x)|^2\ d\mu_X(x)=\int_X |f(\sigmas(x))|^2\ d\mu_X(x)
= \|f\|_{\hilb}^2,
\]
hence $\varphi$ belongs to $L^2(X,\mu_X)$.
Moreover, as $f$ satisfies
condition~{(\ref{condition})}, $F_\sigmas^0\,\varphi=f$,
where now $f$ is regarded as an element of $\hilbm$,
and we conclude that
$\mathrm{Ran}(F_\sigmas^0)=\hilbm$. 
Eventually, notice that, for any $\varphi\in L^2(X,\mu_X)$,
\[
\left(F_\sigmas^0\,\varphi\right)(g)=\chi(\sigmas(\mathsf{p}(g))^{-1}
g)^{-1}\varphi(\mathsf{p}(g))=\left(F_\sigmas\,\varphi\right)(g),\ \ \ \
\forall g\in G.
\]
This completes the proof.~$\blacksquare$

Observe that in the Hilbert space $\hilb$ one can define the
strongly continuous unitary representation $\repres$ of $G$ by
\begin{equation}
\left(\repres_g\,f\right)(g^\prime):= f(g^{-1}g^\prime),\ \ \ \
g,g^\prime\in G,
\end{equation}
for all $f\in\hilb$. Indeed, recalling formula~{(\ref{norm})},
we have:
\begin{eqnarray*}
\|\repres_g\,f\|_{\hilb}^2
\!\! & = & \!\!
\int_G |f(g^{-1}g^\prime)|^2\ d\mu_{G,K}(g^\prime)
\\
\!\! & = & \!\!
\int_G |f(g^\prime)|^2\,\varrho(gg^\prime)\ d\mu_G(g^\prime)
\\
(\ g=\sigmas(x)\,k\ )
\!\! & = & \!\!
\int_{X} |f(\sigmas(x^\prime))|^2\,\zeta(xx^\prime)\
d\mu_X(x^\prime),
\end{eqnarray*}
where
\begin{eqnarray*}
\zeta(xx^\prime)
\!\! & = & \!\!
\int_K \varrho(\sigmas(xx^\prime)(
\kappa_\sigmas(x,x^\prime)^{-1}k_{\sigmas(x^\prime)}k^\prime))\
d\mu_K(k^\prime)
\\
(\,\mbox{invariance of $\mu_K$})
\!\! & = & \!\!
\int_K \varrho(\sigmas(xx^\prime)\, k^\prime)\ d\mu_K(k^\prime).
\end{eqnarray*}
Thus, $\zeta(xx^\prime)=1$ for $\mu_X$-almost all $x^\prime\in X$ and
\[
\|\repres_g\,f\|_{\hilb}^2 =\int_X |f(\sigmas(x^\prime))|^2\ d\mu_X(x^\prime)
= \|f\|_{\hilb}^2.
\]
Then, since $\repres_e=I$ and $\repres_{g_1g_2}=\repres_{g_1}\,\repres_{g_2}$,
for any $g_1,g_2\in G$,
$\repres$ is a unitary representation. Moreover, since,
for any $f,\tilde{f}\in\hilb$, the function
\[
(g,g^\prime)\mapsto f(g^{-1}g^\prime)^*\,\tilde{f}(g^\prime)
\]
is Borel on $G\times G$, we have that
\[
G\ni g\mapsto\langle\repres_g\, f,\tilde{f}\rangle_{\hilb}=
\int_G f(g^{-1}g^\prime)^*\,\tilde{f}(g^\prime)\ d\mu_{G,K}(g^\prime)
\in\mathbb{C}
\]
is a Borel map. This means that $\repres$ is
a weakly Borel unitary representation, hence,
strongly continuous. It follows that
\begin{equation}
\repr :G\ni g\mapsto\repr\!(g)=\isom \repres_g\,\isom^*
\end{equation}
is a strongly continuous unitary representation of $G$ in $\hilbm$.

We want to show now that
the representations $\repr$ and $R^{\chi,\sigmas}$ defined above
can actually be identified. In fact, we can define,
according to Proposition~{\ref{isometry}}, the unitary operator
\begin{equation}
\hat{F}_\sigmas : L^2(X,\mu_X)\ni\varphi\mapsto F_\sigmas\,\varphi\in
\hilbm
\end{equation}
and prove the following result.

\begin{proposition}
The representations $\repr$ and $R^{\chi,\sigmas}$ are unitarily
equivalent. Precisely, we have:
\begin{equation}
\repr\!(g)=\hat{F}_\sigmas\, R^{\chi,\sigmas}_g\,\hat{F}_\sigmas^{*},
\ \ \ \ \forall g\in G.
\end{equation}
\end{proposition}

{\bf Proof\, :} Indeed, for any $f\in\hilbm$, setting
$g^\prime=\sigmas(x^\prime)\,k^\prime$, we have:
\begin{eqnarray*}
\left(\isom^* \unop\,
R_g^{\chi,\sigmas}\,\unop^*\, f \right)\!(\sigmas(x^\prime)\, k^\prime)
\!\! & = & \!\!
\chi(k^\prime)^{-1} \left(\isom^* \unop\, R_g^{\chi,\sigmas}\,\unop^*\,
f\right)\! (\sigmas(x^\prime))
\\
\!\! & = & \!\!
\chi(k^\prime)^{-1} \left(R_g^{\chi,\sigmas}\,\unop^*\, f\right)\!(x^\prime)
\\
\!\! & = & \!\!
\chi(k^\prime)^{-1}\chi(\mathsf{c}_\mathsf{s}(g^{-1},x))
\left(\unop^*\,f\right)\!(g^{-1}[x^\prime])
\\
\!\! & = & \!\!
\chi(k^\prime)^{-1} \chi(\mathsf{c}_\sigmas(g^{-1},x))
\left(\isom^* f\right)\!(\sigmas(g^{-1}[x^\prime]))
\\
\!\! & = & \!\!
\chi({k^\prime}^{-1} \mathsf{c}_\sigmas(g^{-1},x))
\left(\isom^* f\right)\!(g^{-1}\sigmas(x^\prime)\,
\mathsf{c}_\sigmas(g^{-1},x^\prime))
\\ (\ \isom^* f\in \hilb\ )
\!\! & = & \!\!
\left(\isom^* f\right)\!(g^{-1}\sigmas(x^\prime)\, k^\prime)
\\
\!\! & = & \!\!
\left(\repres_g\,\isom^* f\right)\!(\sigmas(x^\prime)\,k^\prime),
\end{eqnarray*}
$\forall g\in G$, $\forall x^\prime\in X$, $\forall k^\prime\in K$.
Then, since $\gamma_\sigmas$ is a bijective map, we have that
\[
\unop\, R_g^{\chi,\sigmas}\,\unop^* = \isom \repres_g\,\isom^*
=\repr\!(g),\ \ \ \ \forall g\in G,
\]
and the proof is complete.~$\blacksquare$

\begin{remark}
The representation $\repres$ in the Hilbert space $\hilb$ is the realization
of the representation of $G$ induced by the representation $\chi$ of $K$
introduced by G.\ Mackey in his seminal paper~\cite{Mackey*} on
induced representations of locally compact groups (extending the work
of Frobenius on representations of finite groups).
This realization is often used in the mathematical literature.
Here we have shown that one can give a realization $\repr$ of this
induced representation in a Hilbert space $\hilbm$ of (equivalence
classes of) functions on $G$ that are square integrable with respect to
a measure in the left-invariant set $\mathcal{M}_{G,K}$, hence
`living completely on $G$', unlike the functions in $\hilb$.
\end{remark}

At this point, we can make the intertwining properties of the
operators $C_\psi^{U,K}$ and $C_\psi^{P_\sigmas}$ explicit.

\begin{theorem}
Let the representation $U$ be square integrable modulo $K$.
Then, for any nonzero vector
$\psi\in\mathcal{A}(U,K)=\mathcal{A}(P_\sigmas)$
the range of the
linear operator $C_\psi^{U,K}$, which is essentially isometric,
is a closed subspace of $\hilbm$ and $C_\psi^{U,K}$ 
intertwines the representation $U$ with the representation $\repr$,
namely
\begin{equation}
C_\psi^{U,K} U(g)= \repr\!(g)\, C_\psi^{U,K},\ \ \ \
\forall g\in G.
\end{equation}
Besides,
the essentially isometric linear operator $C_\psi^{P_\sigmas}$
intertwines  the representation $U$ 
and the projective representation $P_\sigmas$ of $X$ respectively
with the induced representation $R^{\chi,\sigmas}$ and
the left regular $\emme_\sigmas$-representation $R^{\emme_\sigmas}$
of $X$ in $L^2(X,\mu_X)$.
\end{theorem}

{\bf Proof\, :} Observe that, since $K$ is $U$-central, we have:
\[
\langle U(gk)\,\psi,\phi\rangle=\chi(k)^{-1}\langle U(g)\psi,\phi\rangle,
\ \ \ \ \forall g\in G,\ \forall k\in K.
\]
Thus, if $U$ is square integrable modulo $K$, then, for any
$\phi\in\mathcal{H}$ and any $\psi\in\mathcal{A}(U,K)$, $\psi\neq 0$,
the coefficient $c_{\psi,\phi}^{U}$ is a continuous function which
belongs to $L^2(G,\mu_{G,K})$ and
\[
c_{\psi,\phi}^{U}(gk)=\chi(k)^{-1} c_{\psi,\phi}^{U}(g), \ \ \ \
\forall g\in G,\ \forall k\in K.
\]
This proves that $\mathrm{Ran}(C_\psi^{U,K})$ is contained in
$\hilbm$, hence, it is a closed subspace of $\hilbm$ since
$C_{\psi}^{U,K}$ is essentially isometric. Moreover,
\[
\left(C_\psi^{U,K} U(g)\,\phi\right) (g^\prime)=
\langle U(g^\prime)\,\psi, U(g)\phi\rangle = \langle U(g^{-1}g^\prime)\,
\psi,\phi\rangle = c_{\psi,\phi}^U(g^{-1}g^\prime),
\]
hence $C_{\psi}^{U,K}$ intertwines $U$ with $\repr$.\\
Besides, we have already shown in section~{\ref{3}} that
$C_\psi^{P_\sigmas}$ intertwines $P_\sigmas$ with $R^{\emme_\sigmas}$.
Now, let us observe that
\begin{eqnarray*}
\langle P_\sigmas(x)\,\psi,U(g)\,\phi\rangle
\!\! & = & \!\!
\langle U(\sigmas(x))\,\psi,U(g)\phi\rangle
\\
\!\! & = & \!\!
\langle U(g^{-1}\sigmas(x))\,\psi,\phi\rangle
\\
\!\! & = & \!\!
\langle U(\sigmas(g^{-1}[x])\, c_{\sigmas}(g^{-1},x)^{-1})\,\psi,\phi\rangle
\\
\!\! & = & \!\!
\chi(c_\sigmas(g^{-1},x))\,\langle P_\sigmas(g^{-1}[x])\,\psi,\phi\rangle,
\end{eqnarray*}
hence, $C_\psi^{P_\sigmas}$ intertwines $U$ with
$R^{\chi,\sigmas}$.~$\blacksquare$


\section{Discussion of the main results and examples}
\label{6}

We believe that the fundamental points of our paper are two:
\begin{description}

\item[P1]\ \
the definition of square-integrability of a representation $U$ of $G$
modulo a relatively central subgroup $K$ by means of the left-invariant
set of Radon measures $\mathcal{M}_{G,K}$;

\item[P2]\ \
the association of
square integrable representations modulo a relatively central subgroup
with square integrable projective representations.
\end{description}

With respect to the first point, we observe that our definition is a
natural and complete generalization of the usual notion of
square-integrability of a representation. The role played by the Haar
measure $\mu_G$ in the standard case is played, in the generalized case,
by the set $\mathcal{M}_{G,K}$, which consists of measures that are not
left-invariant individually but are transformed one into another by left
translations and reduces (up to normalization) to $\{\mu_G\}$
when $K=\{e\}$. The complete characterization of $\mathcal{M}_{G,K}$
provided by Proposition~\ref{measure} allows to
define, for any measure $\mu_{G,K}$ in $\mathcal{M}_{G,K}$,
the realization $\repr$ in the Hilbert space $\hilbm$ of the
representation of $G$ induced by the 1-dimensional representation $\chi$
of $K$ and to establish the link with square
integrable projective representations, namely the second fundamental
point of our approach. This association has to important consequences.
First, one can prove the `generalized Duflo-Moore'
Theorem~\ref{Duflo3} directly from the classical result of Duflo and Moore,
hence, show that if $U$ is square integrable modulo $K$ then it is
equivalent to a subrepresentation of $\repr$ (that can be regarded as
a natural generalization of the left regular representation of $G$
to which it reduces for $K=\{e\}$).
Second, one can check if the representation $U$ of $G$ is square
integrable modulo $K$ investigating the square-integrability of the
unitary representation $U_{P_\sigmas}$ of the central extension
$X_{\emme_\sigmas}$ of $\mathbb{T}$ by $X=G/K$. 
In many concrete applications, it turns out that $X_{\emme_\sigmas}$
is a semidirect product and one can use some general results
on square integrable representations of semidirect products
(see~{\cite{Aniello}}).

The definition of square-integrability modulo a subgroup of a repesentation
given in this paper can be partially compared with the notion of
representation with a `$\alpha$-admissible'~\cite{Healy} or
`$V$-admissible'~\cite{Ali3} subspace. Anyway, as far as we know, the
points {\bf P1} and {\bf P2} (and their consequences) are specific of
our work.

We will now discuss the example of the representations of the
Weyl-Heisenberg group, which is remarkable for its physical applications,
and an example of a representation that is square integrable modulo
a non-central relatively central subgroup.\\
Let us consider the (2n+1)-dimensional {\it polarized} Weyl-Heisenberg
group (see, for instance,~\cite{Thanga}), namely the semidirect product
\[
\mathbb{H}_n^\prime =
(K\times \Pn)\times_{\act}^\prime \Qn,
\]
where $K$, $\Pn$ and $\Qn$ are vector groups isomorphic
respectively to $\mathbb{R}$, $\mathbb{R}^n$
and $\mathbb{R}^n$, and the action $\act$ of $\Qn$ on
$K\times \Pn$, is defined by:
\begin{equation}
\act_{\mathbf{q}} (k ,\mathbf{p}) =
(k + \mathbf{q}\cdot\mathbf{p},\mathbf{p}),\ \ \
(k,\mathbf{p},\mathbf{q})\in K\times \Pn\times \Qn;
\end{equation}
here the dot denotes the euclidean product.
Thus, the composition law in $\mathbb{H}_n^\prime$ is
given explicitly by
\begin{eqnarray}
(k,\mathbf{p},\mathbf{q})\,(k^\prime,\mathbf{p}^\prime,\mathbf{q}^\prime)
\!\! & = & \!\!
(k+k^\prime+\act_\qqq(k^\prime,\ppp^\prime),\ppp+\ppp^\prime,
\qqq+\qqq^\prime)
\nonumber \\
\!\! & = & \!\!
(k+k^\prime+\mathbf{q}\cdot\mathbf{p}^\prime,
\mathbf{p}+\mathbf{p}^\prime,\mathbf{q}+\mathbf{q}^\prime).
\end{eqnarray}
As the action $\act$ is smooth, $\mathbb{H}_n^\prime$ is a Lie group
(hence, a l.c.s.c.\ group).
The subgroup $K$ is the centre of $\mathbb{H}_n^\prime$.
Then, since $\mathbb{H}_n^\prime$ has a noncompact centre,
according to Proposition~{\ref{compact}}, it cannot admit square integrable
unitary representations. We can also check this result by
explicitly classifying the irreducible unitary representations of
$\mathbb{H}_n^\prime$. In fact, the polarized Weyl-Heisenberg group
has an abelian normal factor $K\times \Pn$, so that we can use
Mackey's little group method.\\
To this aim, let us
identify the dual group $\check{K}\times\check{\Pn}$ of the
normal factor $K\times \Pn$ of $\mathbb{H}_n^\prime$ with
$\mathbb{R}\times\mathbb{R}^n$ by means of the standard pairing:
\begin{eqnarray} \nonumber
(\check{K}\!\times\! \check{\Pn})\!\times\!
(K\!\times\! \Pn)  \cong 
(\mathbb{R}\!\times\!\mathbb{R}^n)\!\times\!
(\mathbb{R}\!\times\!\mathbb{R}^n)
\! & \rightarrow & \!
\mathbb{T} 
\\ \nonumber
\left((\check{k},\check{\mathbf{p}}),
(k,\mathbf{p})\right)
\! & \mapsto & \!
(\check{k},\check{\mathbf{p}})\diamond(k,\mathbf{p})
\equiv e^{i (k\check{k} +
\mathbf{p}\cdot\check{\mathbf{p}})}.
\end{eqnarray}
Then, we have that the dual action $\check{\act}$
of $\Qn$ upon $\check{K}\times\check{\Pn}$,
which is defined by
\[
(\check{\act}_{\mathbf{q}} (\check{k},\check{\mathbf{p}}))\diamond
(k,\mathbf{p}):= (\check{k},\check{\mathbf{p}})\diamond
(\act_{-\mathbf{q}}(k,\mathbf{p})),
\]
has the following explicit form:
\begin{equation}\label{dac}
\check{\act}_{\mathbf{q}}(\check{k},\check{\mathbf{p}}) =
(\check{k},\check{\mathbf{p}} - \check{k}\,\mathbf{q}), \ \ \
(\check{k},\check{\mathbf{p}})\in
\check{K}\times\check{\Pn},\ \mathbf{q}\in \Qn.
\end{equation}
Hence, the $\Qn$-orbits in
$\check{K}\times \check{\Pn}\cong\mathbb{R}\times\mathbb{R}^n$ with
respect to this action can be classified as follows:
\begin{itemize}
\item the singleton orbits
\[
\mathcal{O}_{0,\check{\mathbf{p}}}=\{\,(0,\check{\mathbf{p}})\,\},\ \ \
\check{\mathbf{p}}\in\mathbb{R}^n;
\]
\item the non-singleton orbits
\[
\mathcal{O}_{\check{k}} = \{(\check{k},\check{\mathbf{p}})|\
\check{\mathbf{p}}\in\mathbb{R}^n\},\ \ \
\check{k}\in \mathbb{R}\! - \!\{0\},
\]
which are n-dimensional affine submanifolds of
$\mathbb{R}\times\mathbb{R}^n$.
\end{itemize}
Observe that these orbits are closed subsets of
$\mathbb{R}\times\mathbb{R}^n$.
Then, since $\mathbb{H}_n^\prime$ is a l.c.s.c.\ group,
the orbit structure generated by the dual action $\check{\act}$
is regular (see~{\cite{Folland}}, chapter~6;
or `smooth', see~{\cite{Raja}}, chapter~VI).
It follows that
any irreducible unitary representation of $\mathbb{H}_n^\prime$ is
unitarily equivalent to one generated by Mackey's method.
At this point, since all the orbits generated by $\check{\act}$ are
negligible sets with respect to the
Haar measure $d\check{k}\, d\check{\mathbf{p}}$
on $\check{K}\times\check{\Pn}$, by virtue of a general result concerning
semidirect products with abelian normal factor
(see~\cite{Aniello}, Theorem~2), we
conclude again that $\mathbb{H}_n^\prime$ does not admit
square integrable representations.

We want to show now that the representations associated with the
non-singleton orbits are square integrable modulo $K$. To this aim, let us
consider the generic non-singleton orbit $\mathcal{O}_{\check{k}}$
($\check{k}\in\mathbb{R}\! - \!\{0\}$). The action of $\Qn$ on
$\mathcal{O}_{\check{k}}$ is free, hence, 
there is only one irreducible unitary representation
$U_{\check{k}}$ associated by Mackey's method with
this orbit, i.e.\ the one which is induced by the unitary character
$(\check{k},0)\diamond(\cdot,\cdot)$ of the subgroup $K\times \Pn$
of $\mathbb{H}_n^\prime$. This representation can be realized in the
Hilbert space
$\mathcal{H}_{\check{k}}\equiv L^2(\mathcal{O}_{\check{k}},
d\check{\mathbf{p}})$ and is defined by
\begin{equation} \label{rapp}
\left(U_{\check{k}}(k,\mathbf{p},\mathbf{q})\, f\right)(\mathbf{\check{p}})
=e^{i(k\check{k}+\mathbf{p}\cdot\check{\mathbf{p}})}\,
f(\check{\mathbf{p}}+\check{k}\,\mathbf{q}),\ \ \ \
\forall f\in\mathcal{H}_{\check{k}},
\end{equation}
where we have used the fact that $d\check{\ppp}$ is an invariant measure
with respect to the dual action of $\Qn$. Observe that the restriction
of $U_{\check{k}}$ to $K$ is of the form $\chi_{\check{k}}\,I$, where
$\chi_{\check{k}}$ is the unitary character
$K\ni k\mapsto e^{i k\check{k}}$.\\
Next, one can show easily that the
quotient group $X\equiv\mathbb{H}_n^\prime/K$ can be identified with
the direct product group
$\Pn\times \Qn\cong \mathbb{R}^n\times\mathbb{R}^n$
(notice that this group is not a subgroup of $\mathbb{H}_n^\prime$).
A smooth section $\mathsf{s}:X\rightarrow\mathbb{H}_n^\prime$ is defined by
\begin{equation}
\mathsf{s}(\mathbf{p},\mathbf{q})=(0,\mathbf{p},\mathbf{q}),\ \ \ \
(\ppp,\qqq)\in \Pn\times \Qn.
\end{equation}
Notice, in passing, that another smooth section $\sigmas^\prime$  is given by
\begin{equation} \label{another}
\sigmas^\prime(\ppp,\qqq)=(\ppp\cdot\qqq/2,\ppp,\qqq).
\end{equation}
Then, since
\[
(0,\ppp_1+\ppp_2,\qqq_1+\qqq_2)=
(0,\ppp_1,\qqq_1)\,(0,\ppp_2,\qqq_2)\,(-\qqq_1\cdot\ppp_2,0,0),
\]
we find that
\[
\kappa_\sigmas((\ppp_1,\qqq_1),(\ppp_2,\qqq_2))=
(-\qqq_1\cdot\ppp_2,0,0), \ \ \
\emme_{\check{k},\sigmas}((\ppp_1,\qqq_1),(\ppp_2,\qqq_2))=
e^{-i\check{k}\,\qqq_1\cdot\ppp_2},
\]
where we have used the notations introduced in section~\ref{4}.
Thus, the formula
\begin{equation} \label{proj}
\left(P_{\check{k},\sigmas}(\mathbf{p},\mathbf{q})\, f\right)(\check{\mathbf{p}})
:=\left(U_{\check{k}}(\sigmas(\mathbf{p},\mathbf{q}))\,f\right)
(\check{\mathbf{p}}),\ \ \ \ f\in\mathcal{H}_{\check{k}},
\end{equation}
defines a projective representation of $X$, with multiplier
$\emme_{\check{k},\sigmas}$. According to Proposition~\ref{equivalence},
the unitary representation $U_{\check{k}}$ is square integrable if
and only if $P_{\check{k},\sigmas}$ is. Moreover, as it has been shown in
section~\ref{3},  the problem of the square-integrability of
$P_{\check{k},\sigmas}$ can be tackled by introducing a central extension of
the circle group $\mathbb{T}$ by $X$. In fact, following the recipe
given therein, we define the group $X_{\emme_{\check{k},\sigmas}^*}$
(see Remark~\ref{alternative}) consisting of the cartesian
product $\mathbb{T}\times \Pn\times \Qn$ equipped with the composition
law
\begin{equation}
(\tau,\ppp,\qqq)\,(\tau^\prime,\ppp^\prime,\qqq^\prime)=
(\tau\,\tau^\prime\, e^{i\check{k}\,\qqq\cdot\ppp^\prime}, \ppp+\ppp^\prime,
\qqq+\qqq^\prime).
\end{equation}
$X_{\emme_{\check{k},\sigmas}^*}$, endowed with the product topology,
becomes a l.c.s.c.\ group
(actually, it is even a Lie group).
Notice that $X_{\emme_{\check{k},\sigmas}^*}$ is a semidirect product, i.e.
\begin{equation}
X_{\emme_{\check{k},\sigmas}^*}=(\mathbb{T}\times \Pn)
\times_{\bar{\act}^{\check{k}}}^\prime \Qn,
\end{equation}
where the action $\bar{\act}^{\check{k}}$ of $\Qn$ on $\mathbb{T}\times \Pn$
is defined by
\begin{equation}
\bar{\act}^{\check{k}}_\mathbf{q}(\tau,\ppp)=
(\tau\,e^{i\check{k}\, \qqq\cdot\ppp},\ppp).
\end{equation}
Let us observe explicitly that $X_{\emme_{\check{k},\sigmas}^*}=
X_{\emme_{-\check{k},\sigmas}}$, $\check{k}\in\mathbb{R}\!-\!\{0\}$.
In particular, $X_{\emme_{1,\sigmas}^*}$ is called the {\it reduced polarized
$n$-dimensional Weyl-Heisenberg group}, denoted by
$\bar{\mathbb{H}}_n^\prime$. Now, the dual group of $\mathbb{T}\times \Pn$
can be identified with $\mathbb{Z}\times\check{\Pn}\cong
\mathbb{Z}\times\mathbb{R}^n$ using the standard pairing
\begin{equation}
(\mathbb{Z}\!\times\!\check{\Pn})\!\times\!(\mathbb{T}\!\times\! \Pn)
\ni((j,\check{\ppp}),(\tau,\ppp))\mapsto \tau^j\, e^{i\,\ppp\cdot
\check{\ppp}}.
\end{equation}
Then, since the dual action of $\Qn$ on
$\mathbb{Z}\times\check{\Pn}$ is free,
Mackey's method associates with the generic non-singleton orbit
\[
\mathcal{O}_j=\{(j,\check{\ppp})|\ \check{\ppp}\in\mathbb{R}^n\},\ \ \
j\in\mathbb{Z}\!-\!\{0\},
\]
a single unitary representation $U_{\check{k},j}$
in $L^2(\mathcal{O}_j, d\check{\ppp})$,
defined by
\begin{equation}
\left(U_{\check{k},j}(\tau,\ppp,\qqq)\,f\right)(\check{\ppp})=
\tau^j\, e^{i\,\ppp\cdot\check{\ppp}}\, f(\check{\ppp}
+j\,\check{k}\, \qqq).
\end{equation}
This time the orbit $\mathcal{O}_j$ is a non-negligible set with
respect to the Haar measure $dj\, d\check{\ppp}$ on
$\mathbb{Z}\times \check{\Pn}$, where $dj$ is the counting measure on
$\mathbb{Z}$. It follows that (see~\cite{Aniello}, Corollary~1),
for any $j\in\mathbb{Z}\!-\!\{0\}$, $U_{\check{k},j}$
is square integrable. Now, observe that $U_{\check{k},1}$
coincides with the
representation $U_{*P_{\check{k},\sigmas}}$ associated with the projective
representation $P_{\check{k},\sigmas}$ (see formula~{(\ref{star})}),
or, equivalently, $U_{\check{k},-1}$ with $U_{P_{-\check{k},\sigmas}}$.
Hence, according to Proposition~\ref{equivalence},
for any $\check{k}\in\mathbb{R}\!-\!\{0\}$, the representation
$U_{\check{k}}$ is square integrable modulo $K$.
Moreover, since $X$ is unimodular, the set of the admissible vectors
$\mathcal{A}(U_{\check{k}},K)$
coincides with the whole Hilbert space $\mathcal{H}_{\check{k}}$.

As we have seen in section~{\ref{4}}, the projective representation
$P_{\check{k},\sigmas^\prime}$ associated with the section $\sigmas^\prime$
(see formula~{(\ref{another})}) is
square integrable since $P_{\check{k},\sigmas}$ is.
Now, denoted by $\hat{\qqq},\,
\hat{\ppp}$ respectively the (vector)
position and momentum operators and by
$\hat{\boldsymbol{a}},\,\hat{\boldsymbol{a}}^\dagger$
the (vector) annihilation and creation operators,
\[
\hat{\boldsymbol{a}}=\frac{1}{\sqrt{2}} (\hat{\qqq}+i\,\hat{\ppp}),\ \ \
\hat{\boldsymbol{a}}^\dagger=\frac{1}{\sqrt{2}} (\hat{\qqq}-i\,\hat{\ppp}),
\]
we recall that the displacement operator
with parameter $\para=\frac{1}{\sqrt{2}}(\qqq+i\ppp)$
is defined by
\begin{equation}
\mathcal{D}(\para):=
\exp(\para\!\cdot\!\hat{\boldsymbol{a}}^\dagger-
\para^*\!\!\cdot\! \hat{\boldsymbol{a}}).
\end{equation}
Hence, fixing $\check{k}=-1$ (recall formulae~{(\ref{rapp})},~{(\ref{proj})})
and
identifying $L^2(\mathcal{O}_{-1}, d\check{\ppp})$ with
$L^2(\mathbb{R}^n)$, we have:
\begin{equation}
\mathcal{D}(\para)= e^{i(\ppp\cdot\hat{\qqq}-\qqq\cdot\hat{\ppp})}=
e^{-\frac{i}{2}\,\ppp\cdot\qqq}\, e^{i\,\ppp\cdot\hat{\qqq}}\,
e^{-i\,\qqq\cdot\hat{\ppp}}=P_{-1,\sigmas^\prime}(\qqq,\ppp).
\end{equation}
Thus, the set of displacement operators which generate classical
coherent states (see~\cite{Perelomov})
are nothing but a square integrable projective representation
of $\mathbb{C}$ regarded as a vector group.\\
In many physical applications (for instance, canonical quantization),
the standard Weyl Heisenberg group $\mathbb{H}_n$ is used instead of
$\mathbb{H}_n^\prime$. We recall that $\mathbb{H}_n$ is the Lie group
with manifold $\mathbb{R}\times\mathbb{R}^n\times\mathbb{R}^n$ and
composition law
\begin{equation}
(k,\ppp,\qqq)\,(k^\prime, \ppp^\prime, \qqq^\prime)=
(k+k^\prime+(\qqq\cdot \ppp^\prime - \ppp\cdot \qqq^\prime)/2,
\ppp +\ppp^\prime, \qqq+\qqq^\prime).
\end{equation}
The groups $\mathbb{H}_n$ and $\mathbb{H}_n^\prime$ are isomorphic.
Precisely, the map
\begin{equation}
\iso :\mathbb{H}_n \ni (k,\ppp, \qqq)\mapsto
\left(k + \frac{1}{2}\,\ppp\cdot\qqq, \ppp , \qqq\right)
\in\mathbb{H}_n^\prime
\end{equation}
is an isomorphism of Lie groups.
Here we have used the polarized Weyl-Heisenberg group just because
its semidirect product structure emerges in a more transparent way.\\
We stress also that we have chosen the symbols denoting the elements of
$\mathbb{H}_n^\prime$ so that to obtain the projective
representation which generates the coherent states without using
the Fourier transform. A choice closer to the physicist's point of view
is the following. Interchange the symbols $\ppp$ and $\qqq$ so that, now,
$\mathbb{H}^\prime$ is the semidirect product
$(K\times\mathbf{Q})\times^\prime\mathbf{P}$, with composition law
\begin{equation}
(k,\qqq,\ppp)\,(k^\prime,\qqq^\prime,\ppp^\prime)=
(k+k^\prime+\ppp\cdot\qqq^\prime,\qqq^\prime,\ppp^\prime).
\end{equation}
Next, proceeding as above, we find that the representation associated
by Mackey's method with the nonsingleton orbit $\mathcal{O}_{\check{k}}$
is given by:
\begin{equation}
\left(U_{\check{k}}(k,\qqq,\ppp)\,f\right)(\check{\qqq})=
e^{i(k\check{k}+\qqq\cdot\check{\qqq})}\, f(\check{\qqq}+\check{k}\,\ppp),
\ \ \ \ f\in L^2(\mathcal{O}_{\check{k}},d\check{\qqq}).
\end{equation}
At this point, given the section $\sigmas^\prime:\mathbf{Q}\times\mathbf{P}
\ni (\qqq,\ppp) \mapsto (\qqq\cdot\ppp/2,\qqq,\ppp)\in\mathbb{H}_n^\prime$
and setting $\check{k}=1$,
we have that the equation
\begin{equation}
P(\qqq,\ppp)=U_{1}(\sigmas^\prime(\qqq,\ppp))
\end{equation}
defines a square integrable projective representation $P$.
Then, denoted by $\hat{\mathcal{F}}$ the Fourier-Plancherel operator
in $L^2(\mathbb{R}^n)$,
we have:
\begin{equation}
\hat{\mathcal{F}}\,
\mathcal{D}(\para)\,\hat{\mathcal{F}}^{-1} =  
e^{i(\qqq\cdot\hat{\qqq}+ \ppp\cdot\hat{\ppp})} =
e^{\frac{i}{2}\, \qqq\cdot\ppp}\, e^{i\,\qqq\cdot\hat{\qqq}}\,
e^{i\, \ppp\cdot\hat{\ppp}} = P(\qqq,\ppp),
\end{equation}
where we have identified $L^2(\mathcal{O}_1,d\check{\qqq})$ with
$L^2(\mathbb{R}^n)$. 

We conclude this section with a further example.
Let us consider the group
\[
G=\left(\TTT\times\SSS\times\BBB\times\Pn\right)\times^\prime
\left(\Qn\times(\RRR\times^\prime \AAA)\right),
\]
where
\[
\TTT=\SSS=\BBB=\mathbb{R},\ \ \
\Pn=\Qn=\RRR=\mathbb{R}^n\ \ n\in\mathbb{N},\ \ \
\AAA=\mathbb{R}^+_*,
\]
and the product is defined as follows. First, we specify that
the subgroup $\Qn\times(\RRR\times^\prime\AAA)$ of $G$ is the direct product
of $\Qn$ with the
($n$+1)-dimensional affine group $\RRR\times^\prime\AAA$;
thus, its composition law is given by:
\begin{equation}
(\qqq,\rrr,a)\, (\qqq^\prime,\rrr^\prime,a^\prime)=
(\qqq+\qqq^\prime,\rrr+a\,\rrr^\prime,aa^\prime).
\end{equation}
Next, we define
the action  $(\cdot)[\,\cdot\,]$  of this group on the abelian
normal factor $\TTT\times\SSS\times\BBB\times\Pn$
of $G$ by
\begin{equation}
(\qqq,\rrr,a)[t,s,b,\ppp]=
(t+\qqq\cdot\ppp, a\,s+\rrr\cdot\ppp, a\,b, \ppp).
\end{equation}
At this point,
one can easily check that the map $(\qqq,\rrr,a)\mapsto
(\qqq,\rrr,a)[\,\cdot\,]$ is a homomorphism of
$\Qn\times(\RRR\times^\prime\AAA)$ into the group of automorphisms of
the normal factor of $G$.
Thus, the product in $G$ is well defined
by the action $(\cdot)[\,\cdot\,]$ and is
given explicitly by
\[
(t,s,b,\ppp,\qqq,\rrr,a)\, (t^\prime, s^\prime, b^\prime,
\ppp^\prime,\qqq^\prime,\rrr^\prime,a^\prime)=
(t'', s'', b'', \ppp+\ppp^\prime, \qqq+\qqq^\prime, r'',aa^\prime),
\]
where:
\[
t''=t+t^\prime+\qqq\cdot\ppp^\prime,\
s'' = s+a\,s^\prime+\rrr\cdot\ppp^\prime,\
b''=b+a\, b^\prime,\
\rrr''=\rrr+a\,\rrr^\prime.
\]
Since the action of $\Qn\times(\RRR\times^\prime\AAA)$ on $\TTT\times\SSS
\times\BBB\times\Pn$ is smooth, $G$ is a ($3n$+4)-dimensional Lie group.
Using the pairing
\[
(\check{t},\check{s},\check{b},\check{\ppp})\diamond
(t,s,b,\ppp)= e^{i(t\check{t}+s\check{s}+b\check{b}
+\ppp\cdot\check{\ppp})},
\]
one finds out that the dual action $(\cdot)[\,\cdot\,]\check{\phantom{.}}$
of $\Qn\times(\RRR\times^\prime\AAA)$ on the dual group
$\check{\TTT}\times\check{\SSS}\times\check{\BBB}\times\check{\Pn}
=\mathbb{R}\times\mathbb{R}\times\mathbb{R}\times\mathbb{R}^n$
of the abelian normal factor of $G$ is given by
\begin{equation}
(\qqq,\rrr,a)[\check{t},\check{s},\check{b},
\check{\ppp}]\check{\phantom{.}}=(\check{t},a^{-1}\check{s},
a^{-1}\check{b}, \check{\ppp}-\check{t}\,\qqq-\check{s}\,a^{-1}\rrr).
\end{equation}
Then the orbit $\mathcal{O}_x$, with respect to this action,
passing through the point
$x\equiv(\check{t}_0,\check{s}_0,\check{b}_0,\check{\ppp}_0)=(1,0,1,0)$ of
$\check{\TTT}\times\check{\SSS}\times\check{\BBB}\times\check{\Pn}$
is 
\[
\mathcal{O}_x=\Qn\times(\RRR\times^\prime\Qn) [1,0,1,0]\check{\phantom{.}}
=(1,0,\mathbb{R}^+,\mathbb{R}^n).
\]
The action of $\Qn\times(\RRR\times^\prime\AAA)$ on the orbit $\mathcal{O}_x$
is not free. Indeed, notice that the stabilizer of the point $x$ is the
subgroup $\RRR$.
Hence, according to Mackey's theory, with any unitary character
$\RRR\ni\rrr\mapsto e^{i\,\mathbf{k}\cdot\rrr}$, $\mathbf{k}\in\mathbb{R}^n$,
of $\RRR$ one can associate a unitary representation $U$ of $G$ in
$L^2(\mathcal{O}, d\check{b}\, d\check{\ppp})$ defined by
\begin{equation}
\left(U(t,s,\ldots\!,a)\,f\right)(\check{b},\check{p})=
a^{1/2}\, e^{i(t+\mathbf{k}\cdot\rrr)}\,
e^{i(b\check{b}+\ppp\cdot\check{\ppp}
)}\, F(a\,\check{b},\check{\ppp}+\qqq),
\end{equation}
which is the representation induced by the unitary representation
\[
(\check{t}_0,\check{s}_0,\check{b}_0,\check{\ppp}_0) \diamond
(\cdot,\cdot,\cdot,\cdot)\, e^{i\,\mathbf{k}\cdot(\cdot)}
\]
of the subgroup $(\TTT\times\SSS\times\BBB)\times^\prime\RRR$ of $G$
in the 1-dimensional Hilbert space $\mathbb{C}$.
Then, one can verify easily the following facts:
\begin{itemize}
\item
the direct product group $\TTT\times\SSS\times\RRR$ is a closed normal
subgroup of $G$ and the restriction of $U$ to this subgroup is the
scalar representation $(t,s,\rrr)\mapsto e^{i(t+\mathbf{k}\cdot\rrr)}\,I$,
hence $\TTT\times\SSS\times\RRR$ is $U$-central;

\item
the quotient group $X=G/(\TTT\times\SSS\times\RRR)$ is isomorphic
to the group $(\Pn\times\Qn)\times(\BBB\times^\prime\AAA)$, namely
to the direct product of $\mathbb{R}^{2n}$ with the
(1+1)-dimensional affine group;

\item
a smooth section from $X$ into $G$ is given by
\[
\sigmas :X\ni(\ppp,\qqq,b,a)\mapsto(0,0,b,\ppp,\qqq,0,a)\in G ;
\]

\item
with this section is associated the projective representation
$P_{\sigmas}=U\circ\sigmas$ of $X$, with multiplier $\emme_{\sigmas}$
defined by
\[
\emme_\sigmas((\ppp,\qqq,b,a),(\ppp^\prime,\qqq^\prime,
b^\prime,a^\prime))
= e^{-i\,\qqq\cdot\ppp^\prime};
\]

\item
the projective representation $P_\sigmas$ is square integrable, as
one can check defining the group $X_{\emme_\sigmas}$ and studying its
irreducible unitary representations, or, alternatively, observing that
$P_\sigmas$ is the tensor product of a square integrable projective
representation of $\Pn\times\Qn$ and a square integrable unitary
representation of the (1+1)-dimensional affine group $\BBB\times^\prime\AAA$;

\item
by the previuos item, the irreducible unitary representation $U$ is
square integrable modulo the relatively central
subgroup $\TTT\times\SSS\times\RRR$.

\end{itemize}

Observe that the $U$-central subgroup $\TTT\times\SSS\times\RRR$ is
not central in $G$. Indeed, for instance, for $a\neq 1$, we have:
\begin{eqnarray*}
(0,0,\ldots,0,a)\,(t,s,0,0,0,\rrr,1)\,(0,0,\ldots,a^{-1})
\!\! & = & \!\!
(t,a\,s,0,0,0,a\,\rrr,1)
\\
\!\! & \neq & \!\!
(t,s,0,0,0,\rrr,1).
\end{eqnarray*}
Besides, $X$ is not unimodular. In fact, one finds easily that the
modular function $\Delta_X$ on $X$ is given by
\begin{equation}
\Delta_X(\ppp,\qqq,b,a)=a^{-1},\ \ \ \ \forall(\ppp,\qqq,b,a)\in X.
\end{equation}
Hence, according to Theorem~\ref{Duflo3},  the Duflo-Moore operator
$D_{U,\TTT\times\SSS\times\RRR}$ is not bounded.
Applying Corollary~2 of ref.~\cite{Aniello} to the square integrable
representation $U_{P_\sigmas}$ of $X_{\emme_\sigmas}$,
one finds easily that, with a suitable normalization of the left Haar measure
on $X$, $D_{U,\TTT\times\SSS\times\RRR}$ is the multiplication
operator in $L^2(\mathcal{O}, d\check{\ppp}\,d\check{b})$ by the
function $(\check{\ppp},\check{b})\mapsto\check{b}^{-1/2}$.



\end{document}